\documentclass[final,5p,times]{elsarticle} 
\usepackage{epsfig}  
\usepackage{amssymb,amsmath}
\usepackage{tabularx}
\usepackage[table,xcdraw, dvipsnames]{xcolor}
\usepackage{natbib}
\usepackage[greek,english]{babel}
\usepackage{subcaption}
\usepackage{float}
\usepackage[colorlinks=true,linkcolor=blue,urlcolor=blue]{hyperref}
\usepackage{ulem}
\usepackage{microtype}

\usepackage{lineno}

\begin{document}

\journal{Ecological Modeling} 

\begin{frontmatter} 
\title{Energy, space and competition: A model of territorial organization in camelids}

\date{\today}

\author[tomas]{Tomás Ignacio González\corref{cor1}}
\ead{tomignaciogon@gmail.com}
\cortext[cor1]{Corresponding author}

\author[guillermo]{Guillermo Abramson}
\ead{abramson@cab.cnea.gov.ar}

\author[fabiana]{María Fabiana Laguna}
\ead{lagunaf@cab.cnea.gov.ar}

\address[tomas]{Statistical and Interdisciplinary Physics Division, Centro Atómico Bariloche (CNEA), CONICET and Postgraduate Program in Biology, Universidad Nacional del Comahue. 
R8402AGP Bariloche, Argentina}

\address[guillermo]{Statistical and Interdisciplinary Physics Division, Centro Atómico Bariloche (CNEA), CONICET and Instituto Balseiro (Universidad Nacional de Cuyo). R8402AGP Bariloche, Argentina.}

\address[fabiana]{Statistical and Interdisciplinary Physics Division, Centro Atómico Bariloche (CNEA), CONICET and Universidad Nacional de Río Negro. R8402AGP Bariloche, Argentina.}

\begin{abstract}
 
Ecological theory predicts that herbivores adopt behaviors that balance energetic costs and benefits, shaping both foraging and reproductive organization.  In some wild ungulates, such as camelids, males defend resource-rich areas to attract females and secure their offspring, leading to the emergence of territorial systems. However, existing territorial models rarely incorporate the well-established allometric relationships between body mass and metabolic costs. 

In this study, we develop a spatially explicit, individual-based model of territorial organization in camelids, in which male behavior is governed by an energy balance equation that includes allometric locomotion costs, resource acquisition, and the energetic risks of aggressive encounters. We show that (i) a body-mass-dependent threshold determines which individuals can establish and maintain territories, (ii) spatial resource connectivity strongly modulates competition intensity and territorial inequality, and (iii) confrontation costs play a key stabilizing role by limiting territorial expansion and promoting coexistence. 

Our results provide a mechanistic link between metabolic scaling, spatial structure, and social organization, offering new insights into the ecological and evolutionary drivers of territoriality in ungulates. 

\end{abstract}

\begin{keyword}
territory, energy, males, competition, mathematical modeling, camelids 
\end{keyword}

\end{frontmatter}

\section{Introduction}
Foraging is a fundamental activity in the life cycle of herbivorous animals. To meet their metabolic demands while performing activities such as mating, resource defense, and predator avoidance, individuals must acquire nutrients and energy at sufficient rates. At the same time, they must allocate time and energy among these competing activities in ways that optimize their fitness \citep{herbers1981time, vila1994time, pyke2019optimal, abramson1a2014space}.

Since time and energy allocation must maintain a positive energetic balance while satisfying ecological constraints, foraging behavior is widely assumed to be shaped by natural selection. This idea led to the formulation of Optimal Foraging Theory and to a large body of mathematical models addressing the energetic economy of foragers \citep{pyke2019optimal, werner1981optimal, krebs1984optimization, mitchell2012foraging, kazimierski2016movement}.

Traits shaped by natural selection range from basic ones such as diet, habitat selection, and foraging behavior to more complex patterns such as home range, habitat use, group behavior, and territoriality \citep{vila1994time, krebs1984optimization, giuggioli2006theory, mitchell2012foraging, dill1978energy, smith2020and}. While the home range is defined as the area used by an individual during its life cycle, a territory is the portion of that area from which at least some conspecifics are excluded \citep{mitchell2012foraging, burt1943territoriality}.

In this context, foraging strategies are often closely intertwined with reproductive organization in species such as ungulates. The availability of key resources such as food and water can be used by males to attract one or multiple females. Several classifications of territorial systems have been proposed, including pair territories (defended by a monogamous pair), polygynous territories (defended by a single male with multiple long-term mates), and lek territories (occupied by multiple males with short-term mating opportunities). Territoriality can also be categorized as resource-defense polygyny, where males control access to females through resources, or female-defense polygyny, where males control access to females directly \citep{emlenecology, bowyer2020evolution}.

American wild camelids, particularly \textit{Vicugna vicugna} (vi\-cu\-ñas) and \textit{Lama guanicoe} (guanacos), provide a clear example of resource-based territorial organization. In these species, females are attracted to feeding areas monopolized by territorial males that form family groups. However, the two species differ in their degree of social flexibility: guanacos appear to be more flexible than vicuñas, showing less rigid territoriality, group size, composition, and social hierarchies \citep{lucherini1996aggressive, bravo2001order, marino2014ecological, acebes2018vicuna}.

Another important difference concerns the persistence of territories. In general, vicuñas defend permanent territories, whereas guanaco populations may be either migratory, occupying territories during the mating and warm seasons and abandoning them during the cold season, or sedentary, maintaining territories year-round \citep{franklin1983contrasting, cassini2009sociality, marino2012indirect}.

Males, which typically act as territorial holders, must additionally invest time and energy in defensive tasks such as vigilance, patrolling, and reinforcing territory ownership. These activities may include energetically costly aggressive encounters with intruding males. Such costs may also be influenced by the presence of interspecific competitors and predators \citep{lucherini1996aggressive, cassini2009sociality, flores2020modelling}.

The persistence and size of territories are closely related to resource dynamics. Permanent territories are thought to be advantageous in terms of foraging time only when food resources remain stable and predictable over time. Conversely, territories tend to be larger in spatially variable environments than in more constant ones \citep{dill1978energy, emlenecology, horn1968adaptive, mcnair1987effect}.

These patterns ultimately reflect the energetic demands and costs faced by individuals throughout their life cycle. Camelids spend a large fraction of the day feeding, and this tendency increases during the cold season, when animals must consume and digest low-quality grasses for longer periods in order to meet their energetic and nutritional requirements. Females must also face the energetic costs of gestation and lactation following the mating season \citep{cassini2009sociality, cock2009south, flores2020modelling}.

Because territoriality has an inherently spatial character, it has long attracted interest from theoretical ecologists and mathematical modelers. Over the years, a variety of models have been proposed that consider factors such as food distribution, group size effects, antipredator behavior, and energetic costs \citep{dill1978energy, hinsch2010defence, diaz2001territorial}.

Particularly relevant are the so-called economic models of territoriality, which evaluate optimal balances between benefits and costs under the assumption that behaviors affecting an animal’s fitness have been shaped by natural selection, thus linking territorial behavior to Optimal Foraging Theory \citep{mitchell2012foraging, emlen1966role, macarthur1966optimal, rapport1977economic}. Despite these advances, existing models rarely incorporate the well-established allometric relationships between body mass and energetic expenditure.

Taylor and Heglund \cite{taylor1982energetics} proposed a general explanation for how muscles use metabolic energy during locomotion. Using data from 73 avian and mammalian species, they derived a single equation predicting locomotion cost as a function of body mass. This relationship implies that larger animals incur higher energetic costs during locomotion. Patterns consistent with this scaling have been observed in wild guanaco populations, where larger individuals must face substantial physiological costs associated with activities such as year-round territorial defense. Conversely, smaller body size may represent a selective advantage for males facing reduced foraging requirements. This interpretation is also consistent with observations that guanacos spend more time foraging than vicuñas, since the former are typically much larger than the latter \citep{marino2012indirect, lucherini1996group, raedeke1979population}.

At the same time, dominance and access to females in wild camelids are usually determined through aggressive interactions among males. Traits associated with fighting ability, such as age and large body size, may therefore confer a selective advantage, a pattern widely reported in these species \citep{bowyer2020evolution, lucherini1996aggressive, vanpe2009access}.

To the best of our knowledge, despite the importance of these energetic and behavioral mechanisms, the interaction between body-mass-dependent energetic costs, competitive interactions, and spatial resource structure has not yet been explored within a unified modeling framework for territorial systems.

In this work we develop a spatially explicit mathematical model of the territorial organization of male camelids with different body masses and explore its behavior through numerical simulations. The model incorporates allometric relationships between body mass and locomotion metabolic cost, as well as energetic costs associated with competition, territorial vigilance, and spatial resource distribution. By linking individual energy balance with spatial competition, the model provides a mechanistic framework for understanding how metabolic scaling can influence territorial organization.

\section{Model description}

\subsection{Overview}

\subsubsection*{Purpose and patterns}

 We developed a spatially explicit simulation model to investigate territory formation and competition among a fixed number of camelid males. The number of iterations represents a single season of territoriality, and each iteration could be considered approximately one day. After the simulation, we analyzed the number, size, distribution and shape of the resulting territories to characterize the levels and types of competition under different sets of parameters. As mentioned in the previous section, a territory has strategic value for male reproduction; therefore, territory formation occurs before breeding. Consequently, we do not consider reproductive processes in our simulation.
 
 To characterize the distribution of territories among males, we analyzed the inequality of territorial areas using the Gini index (Fig.~\ref{fig:Estab}B). Originally introduced as a measure of statistical dispersion in economics, the Gini index can be adapted to our system as
 \begin{equation}
    G(t) =\frac{1}{2MA} \sum_{i,j}\left| a_i(t)-a_j(t) \right|,
    \label{eq:g2}
\end{equation}
where $M$ is the number of males, $A$ is the total occupied area of the system, and $a_i(t)$ denotes the area of the territory held by male $i$ at time $t$. Values close to $0$ indicate a more homogeneous distribution of territory sizes, whereas values close to $1$ correspond to highly unequal territorial areas \citep{Farris2010}.

Regarding the shape of the territories, we quantified border complexity through the calculation of the Perimeter-Area Fractal dimension (PAFRAC) proposed by McGarigal et al. (2002) \citep{fragstats}. 

\begin{equation}
\ln(A) = b_1 \ln(L) + b_0, \quad \text{with} \quad PAFRAC = \frac{2}{b_1}.
\label{eq:pafrac}
\end{equation}

For each territory, we computed the log-log regression of area $A$ versus perimeter $L$, and then estimated the index based on the slope of the regression. PAFRAC values close to 1 indicate compact, simple boundaries, whereas higher values of the index indicate greater boundary complexity.

\subsubsection*{Entities, state variables and scales}

The model is based on a $50\times50$ grid representing the habitat, with spatially distributed resources, in which $M$ camelid males attempt to establish their territories. We assume a spatial resolution of 1 ha per grid cell, consistent with reported territory sizes for guanacos and vicuñas \citep{young2004activity, young2004territorial, franklin1974social}.

We assume that all males are identical except for their body mass, which is drawn from a uniform distribution ranging from 50 to 140 kg \citep{comportamiento_vicuna}:
\[
m_i \sim U(50,140), \quad \forall i .
\]

At an earlier stage of model development, with the aim of defining the distribution of resources available to males, we introduced a ``sowing'' model, which is described in more detail in the Submodels section and in the Supplementary Material.

\subsubsection*{Process overview} 

Each male attempts to occupy and acquire cells in that contain resources in order to establish its territory. However, as emphasized by \citep{cassini2009sociality, young2004territorial}, males in the wild must invest time and energy in patrolling their territory boundaries and defending them against intruders. Accordingly, in our model, males must weigh the benefits of continuing territorial expansion against the costs of confronting other males.

A schematic overview of the scheduling processes involved in territorial expansion is provided in Figure~\ref{fig:DTree}, and a detailed description is provided below.
\subsection{Design concepts} 

Territorial expansion is constrained by an energetic trade-off between the benefits of acquiring additional resources and the costs of patrolling and defending the territory, including the energetic costs of locomotion. To quantify locomotion cost, we adopt the empirical allometric relation of Taylor and Heglund \citep{taylor1982energetics}, which estimates the the energetic cost of, $E$, as a function of body mass ($m$) and locomotion speed ($v$):
\begin{equation}
E(m,v) = 10.7 m^{-0.316}v + 6.03 m^{-0.303},
\label{eq:TyH}
\end{equation}
where the result is expressed in W kg$^{-1}$. We consider this equation appropriate for our camelid species because the empirical relationship was derived from data including quadrupedal locomotion of big mass mammals, such as that of horses. 

We further define the energetic balance of each male $i$, $B_i$, as
\begin{equation}
B_i(R_i,m_i,v,P_i) = \mu R_i - m_iE(m_i,v)\frac{P_i}{v},
\label{eq:Balance1}
\end{equation}
where $R_i$ is the total amount of resources contained in the territory of male $i$, and $P_i$ is its perimeter. The second term represents the energetic cost of traversing the territory boundary once at speed $v$.

Based on reported estimates of camelid walking speed \citep{van2010comparative}, we assume a constant value of $v = 1$~m/s for all animals. The parameter $\mu$ is also common to all males and represents the energetic intake per unit resource.

At each iteration of the model, each male evaluates the neighboring cells of its current territory. A cell may be added to the territory if doing so results in a positive energetic gain.

If the cell is unoccupied, it is incorporated when the energetic gain from the additional resources exceeds the cost associated with the additional patrolling required by the enlarged territory perimeter (Eq.~\ref{eq:Balance1}).

If the cell is already occupied, the situation may lead to a confrontation between the intruding male and the current owner. Empirical studies indicate that both residents and intruders can display varying levels of aggression, ranging from no reaction to costly and potentially dangerous fights. This suggests that males may assess the net benefit of engaging in a confrontation associated with acquiring a given resource site \citep{lucherini1996aggressive, geist1971mountain, wilson1985male}.

In our model, when a male evaluates annexing a cell that already belongs to another male's territory, it compares the expected energetic gain with the expected cost associated with losing the confrontation. We assume that males can assess their probability of winning based on their own and their opponent's body mass. Let $\delta = (m_i-m_j)/(m_i+m_j)$ denote the relative  difference in body mass between the intruding male $i$ and the resident male $j$. For arbitrary masses, $\delta\in(-1,1)$. However, such extreme differences are unrealistic in natural populations. For guanacos, documented adult males typically range between 50 and 140~kg, which corresponds to $\delta$ values ranging approximately from $-0.47$ to $0.47$. We therefore consider the slightly broader relevant interval $\delta\in(-1/2,1/2)$. Within this range, we define the probability that the intruding male $i$ defeats the resident male $j$ as:
\begin{equation}
p_{ij} =
\begin{cases}
0 & \text{if } \delta<-1/2, \\
\frac{1}{2} + \delta & \text{if } \delta\in(-1/2,1/2), \\
1 & \text{if } \delta>1/2 .
\end{cases}
\label{eq:Prob}
\end{equation}
This linear form provides a simple approximation that preserves the symmetry condition $p_{ji}=1-p_{ij}$ and ensures a continuous dependence of winning probability on relative body mass. Further details on the choice of this definition, including its relationship with the distribution of body masses and possible generalizations, are provided in the Supplementary Material.

As discussed above, the intruding male initiates the encounter only if the expected energetic payoff of the confrontation is positive, namely if
\begin{equation}
p_{ij}\Delta B_i > (1-p_{ij})C ,
\label{eq:Compare}
\end{equation}
where $\Delta B_i$ is the expected net change in energetic balance associated with acquiring the contested cell:
\[
\Delta B_i = \mu \Delta R_i - m_iE(m_i,v)\frac{\Delta P_i}{v},
\label{eq:DB}
\]
and the parameter $C$ represents the energetic cost or risk incurred if the intruding male loses the encounter. In the simulations, $C$ is treated as a control parameter and varied to explore how the cost of conflict affects the resulting territorial patterns. If an encounter occurs, its outcome is determined stochastically: the intruding male wins the contest for the cell with probability $p_{ij}$.

Finally, if a male loses all cells belonging to its territory, it loses territorial status and is removed from the territorial system. This does not imply the death of the male; rather, it represents a transition to a non-territorial category, corresponding to males that are not associated with females and are typically found either as solitary individuals or in bachelor groups with no associated territory \citep{ lucherini1996aggressive, cassini2009sociality, vila1995spacing}.

\begin{figure}[t]
\centering
\includegraphics[width=\columnwidth]{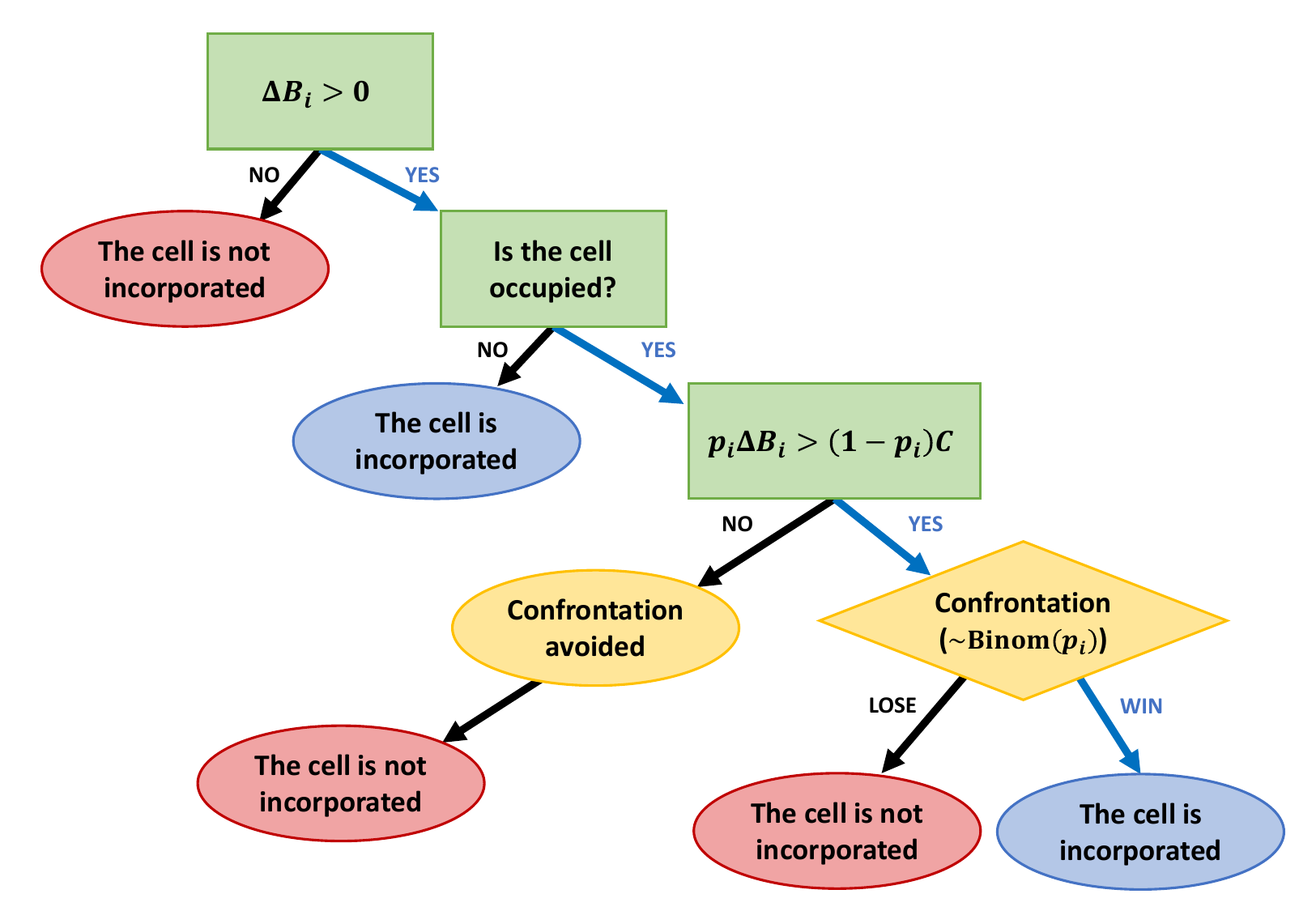}
\caption{Decision tree considering all the parameters, factors and processes that each male must face at each step of the simulation.}
\label{fig:DTree}
\end{figure}

We refer the reader to Figure~\ref{fig:DTree} for a schematic summary of the expansion process of a male's territory through a decision tree, considering both deterministic and stochastic aspects.

\subsection{Details}

\subsubsection*{Initialization}

Initially, once the resource matrix has been defined, each male occupies a single cell and attempts to expand its territory by incorporating neighboring cells containing different levels of resources. 

As a consequence of this random initialization, some individuals may start in cells with very low resource availability, leading to a negative initial energetic balance and causing them to disappear immediately, before the territorial dynamic begins. To avoid this artificial effect, we introduce a rule that applies only during the first iteration of the simulation. Males whose initial territory consists of a single low-resource cell and yields a negative energetic balance are allowed to move to a nearby cell that increases their energetic balance. This displacement is not arbitrary. The maximum distance that a male can travel is constrained by the same energetic balance condition, including the locomotion cost described by the Taylor and Heglund relation \citep{taylor1982energetics}.

The parameters explored in the simulations are the spatial range of resources over the grid, controlled by $\alpha$, the energetic intake provided by resources ($\mu$), and the energetic cost of aggressive encounters ($C$). We consider values of $\alpha\in[0,1]$, values of $\mu$ between 50 and 185~kJ, and C ranging from 0 to 185 kJ.

\subsubsection*{Submodels}

As described above, we use a ``sowing'' model to define the distribution of resource on the grid. This model is defined by two parameters: $N$ sowing points of maximum resource richness, placed at random locations in an initially empty matrix, and $\alpha$, a control parameter ranging from 0 to 1 that determines the spatial propagation of resources from these sowing points.

For all our simulations, the resource matrix is generated using $N=20$ sowing points, while the parameter $\alpha$ is varied to explore different degrees of spatial heterogeneity.

For small values of $\alpha$, resources remain concentrated around the sowing points, and most of the matrix is inhospitable. In contrast, large values of $\alpha$ produce a relatively homogeneous matrix in which  cells have similar resource levels. Intermediate values generate heterogeneous landscapes with varying resource levels across regions of the grid. A full description of this function is provided in the Supplementary Material.

\section{Results}

In this section we analyze how the main parameters of the model affect the final state and stability of the system. The outcomes are characterized by the distribution of territorial areas and energetic balances, as well as by the proportion of persistent (non-excluded) males.

Additionally, we examine how the scarcity of energy resources defines a clear analytic threshold that determines which males are able to establish and maintain a territory.

\subsection{Male persistence and territorial distribution}

\begin{figure}[t]
\centering
\includegraphics[width=0.9\columnwidth]{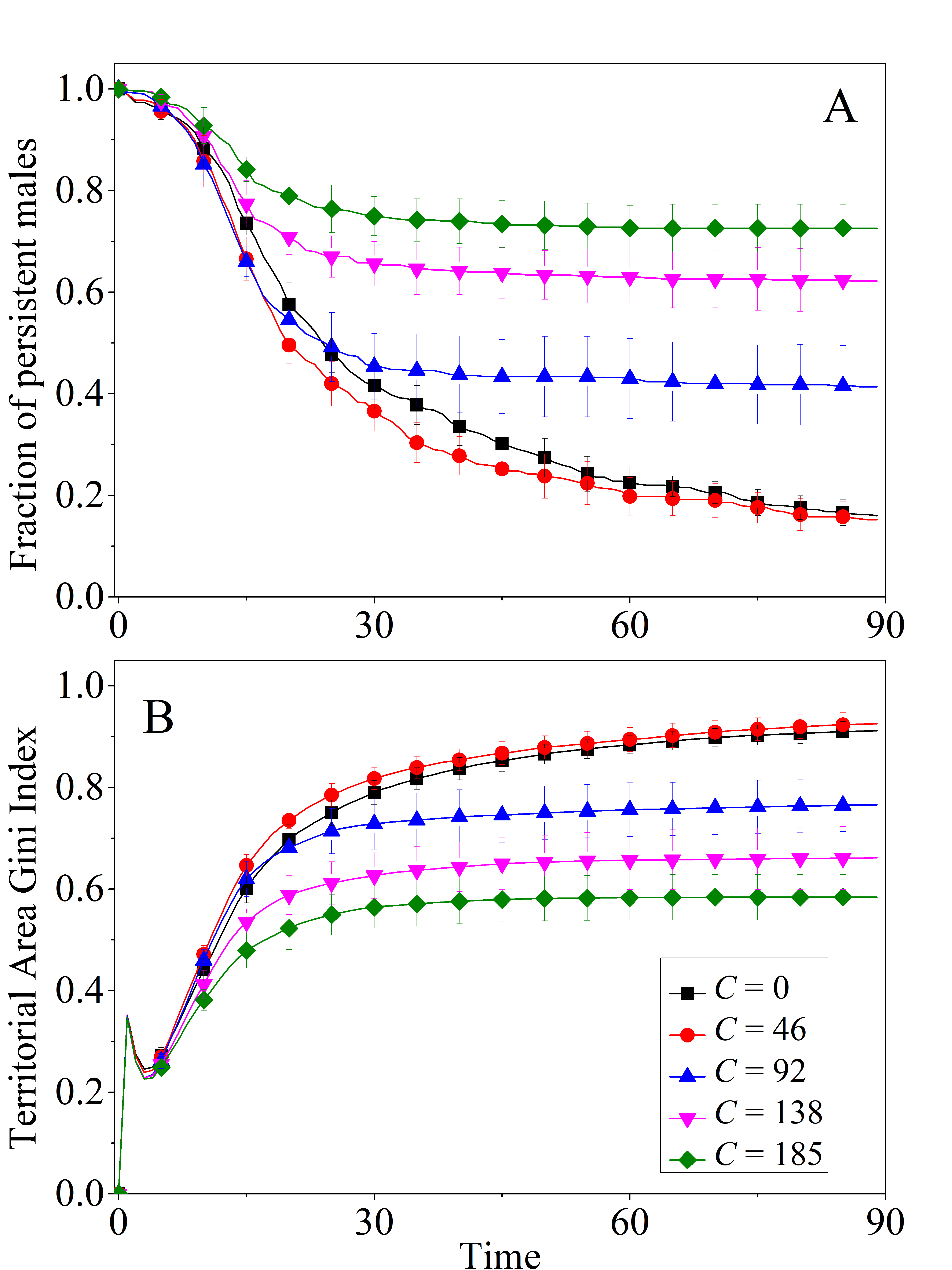}
\caption{A) Fraction of persistent males over time for different levels of confrontation cost $C$.  
B) Evolution of the Gini index of territorial areas for the same cases. Each point represents the average of 10 realizations. For these examples, we used an energy parameter of $\mu = 150$ and a matrix connectivity parameter of $\alpha = 1$.}
\label{fig:Estab}
\end{figure}

We begin by analyzing the temporal evolution of the system as a function of the confrontation cost, $C$. Figure~\ref{fig:Estab} shows the dynamics of two key quantities for different values of this parameter: the fraction of persistent males (A) and the inequality in the distribution of territorial areas, quantified by the Gini index (B). The results correspond to simulations with fixed resource parameters ($\mu = 150$, $\alpha = 1$).

As shown in Fig.~\ref{fig:Estab}A, when there is no confrontation cost ($C=0$, black curve) or when this cost is very small, the fraction of persistent males continues to decrease and does not reach an apparent steady state within the simulated time. In contrast, increasing the confrontation cost stabilizes the system and allows a larger fraction of males to persist.

As shown in Fig.~\ref{fig:Estab}B, the Gini index always reaches a stable value, even in cases where the fraction of persistent males does not stabilize. However, increasing confrontation costs lead to a more homogeneous distribution of territories, reflected in smaller values of the Gini index.

These results show that confrontation costs can stabilize the system by limiting the intensity of territorial conflicts. However, stabilization can also emerge from spatial constraints on resource distribution, as shown below.

\begin{figure}[t]
\centering
\includegraphics[width=1.1\columnwidth]{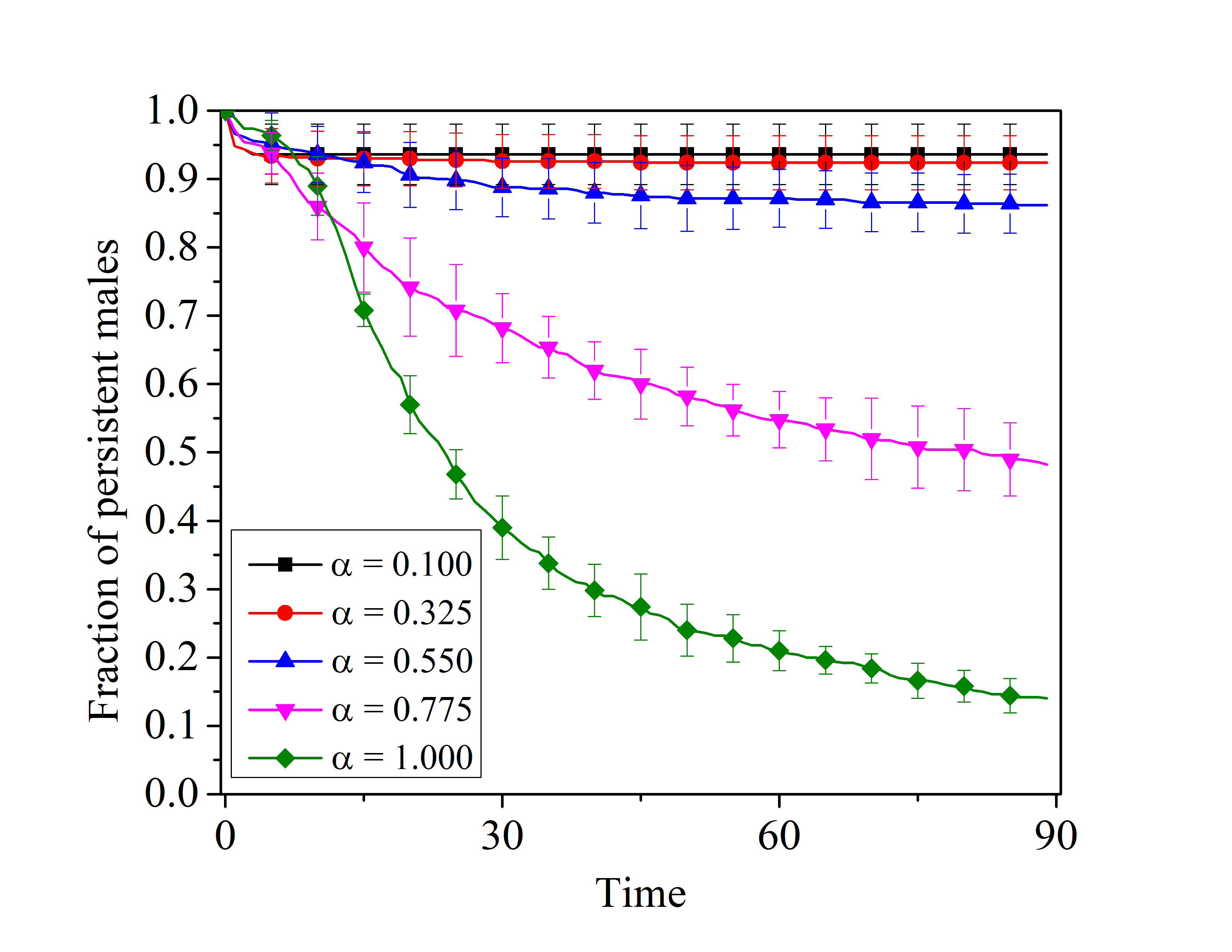}
\caption{Fraction of persistent males over time for different levels of matrix connectivity $\alpha$. Each curve represents the average of 10 realizations. For these examples, we fixed the energy parameter $\mu = 150$ and the confrontation cost $C = 0$.}
\label{fig:Estab2}
\end{figure}

An alternative way to stabilize the system, without increasing the confrontation cost, is by using a resource matrix with limited propagation. The results shown in Fig.~\ref{fig:Estab2} are indicative that smaller values of $\alpha$ not only lead to stabilization but also result in higher fractions of persistent males. In contrast, higher values of $\alpha$ produce decreasing curves that fail to stabilize, and almost all males are eventually eliminated through competition.

\begin{figure}[h]
\centering
\includegraphics[width=\columnwidth]{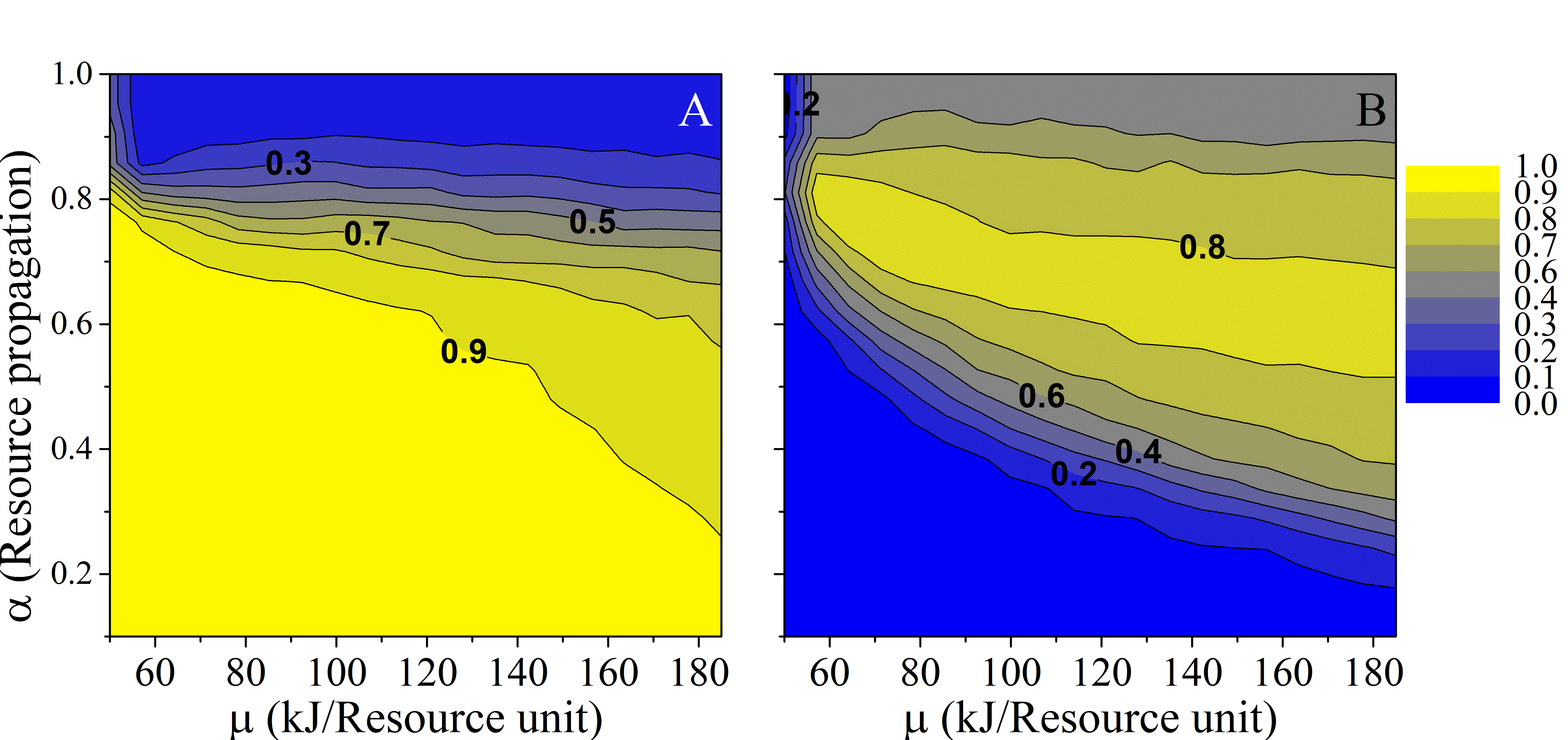}
\caption{Heatmaps in the $(\alpha, \mu)$ plane. Panel A: fraction of persistent males after 90 iterations; Panel B: Gini index of territorial areas. Each pixel represents the average over 20 realizations. The confrontation cost was fixed at $C = 0$ for all simulations.}
\label{fig:AlphaMu}
\end{figure}

We also explored how the final state of the system depends simultaneously on resource connectivity and energetic richness by computing the fraction of persistent males and the distribution of territorial areas across the $(\alpha, \mu)$ parameter plane. The results are shown as heatmaps in Fig.~\ref{fig:AlphaMu}, with the fraction of persistent males in Fig.~\ref{fig:AlphaMu}A and the Gini index of territorial areas in Fig.~\ref{fig:AlphaMu}B.

For low values of $\alpha$ (poorly connected resources) or low $\mu$ (energetically limited resources), most males persist in the system (Fig.~\ref{fig:AlphaMu}A). However, this is not the result of competitive equilibrium. Males rarely encounter each other and remain in their initial territories, which retain minimal size. This is reflected in Fig.~\ref{fig:AlphaMu}B, where the Gini index is low, indicating a nearly uniform distribution of territory sizes. 

Increasing either $\alpha$ or $\mu$ leads to stronger competition: fewer males can maintain territories, and the fraction of persistent males decreases nonlinearly as the system becomes more homogeneous and energetically rich (Fig.~\ref{fig:AlphaMu}A). The Gini index exhibits a non-monotonic response, meaning that an increase in both $\alpha$ and $\mu$ initially increases the Gini index, but it subsequently decreases as these parameters are further increased (Fig.~\ref{fig:AlphaMu}B). Territories are initially evenly distributed when resources are poor and localized, then become more uneven (higher Gini) as either parameter increases, reaching a maximum in an intermediate region. This indicates that some males monopolize large resource patches while others are restricted to smaller, disconnected areas. Finally, when resources are abundant and fully connected, competition becomes widespread and territories are redistributed more evenly, reducing the Gini index again (Fig.~\ref{fig:AlphaMu}B).

We also examined how population and grid size influence the dynamics of the model. These two parameters have opposite effects: increasing the grid size promotes male persistence but reduces variability in territorial areas, whereas increasing the population size decreases male persistence and increases the inequality of territory sizes. These results are shown in the Supplementary Material.

\subsection{Final mass distribution}

The relationship between body mass and energy requirements, i.e., the energetic balance equation (Eq.~\ref{eq:Balance1}), defines a mass threshold that determines how large a male can be, while still holding a territory and maintaining at least a null balance ($B_i = 0$). This threshold is given by:
\begin{equation}
m^* < \psi \mu^{1.45},
    \label{eq:Threshold}
\end{equation}
where $\psi$ is a numerical coefficient obtained from approximations that combines the spatial scale and shape defining the territory. It also incorporates the allometric relationships between body mass and energy costs derived from the Taylor and Heglund equation (Eq.~\ref{eq:TyH}). To determine the threshold condition, we treat $\psi$ as a constant and set $\psi = 7.75 \times 10^{-6}~\mbox{J}^{-1.45}\mbox{kg}$, corresponding to the initial territorial configuration of one square cell. The derivation of this value is provided in the Supplementary Material.

\begin{figure}[t]
\centering
\includegraphics[width=1.1\columnwidth]{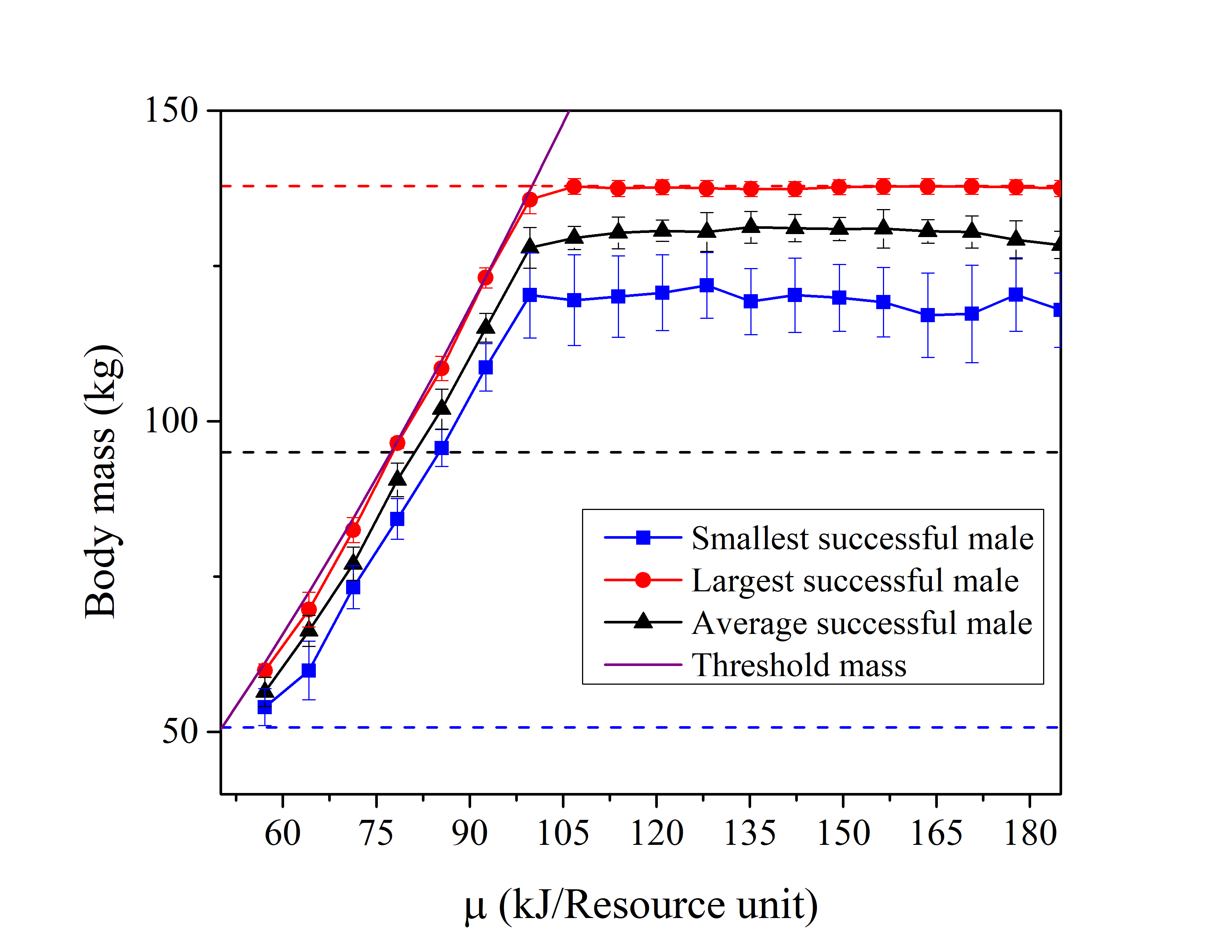}
\caption{Minimum (blue), average (black), and maximum (red) values of the final mass distribution of males that were able to establish a territory of size greater than 1 (initial condition), in relation with the energy available in the resource, $\mu$. The purple curve is the threshold given by Eq.~(\ref{eq:Threshold}). Each curve is the average of 10 realizations. To isolate the effect of $\mu$, we fixed the parameters $\alpha = 1$ and $C = 0$. The dashed lines indicate the average minimum (blue), mean (black) and maximum (red) values of the initial mass distribution, considering a total of 50 males. }
\label{fig:Masa}
\end{figure}

To evaluate the threshold condition, we performed a series of simulations with 50 males of varying body mass, and recorded the masses of the largest and smallest persistent males retaining territories larger than one cell. For these simulations, we fixed the sowing dispersion parameter $\alpha = 1$, corresponding to a fully connected and homogeneous resource distribution, and set the confrontation cost $C = 0$, meaning that males do not expend energy on aggressive interactions. These choices isolate the effect of the energetic input $\mu$ on male persistence and mass distribution.

As shown in Fig.~\ref{fig:Masa}, increasing the energetic intake provided by resources ($\mu$) has a direct effect on the final mass distribution of males at the end of the simulations.
For very small values of $\mu$, when even the smallest males fail to satisfy the threshold condition (Eq.~\ref{eq:Threshold}), no individual can establish a territory larger than one cell.
As $\mu$ increases, the maximum value of the final mass distribution (Fig.~\ref{fig:Masa}, red curve) rises rapidly and reaches the upper bound of the initial distribution, indicating that even the largest males successfully establish territories at intermediate parameter values.
A similar trend is observed for the minimum mass (Fig.~\ref{fig:Masa},  blue curve), which increases with $\mu$ and eventually saturates above the mean of the initial distribution, indicating that smaller males are excluded from the system. Beyond this point, no further increase is observed.

\begin{figure}[t]
\centering
\includegraphics[width=\columnwidth]{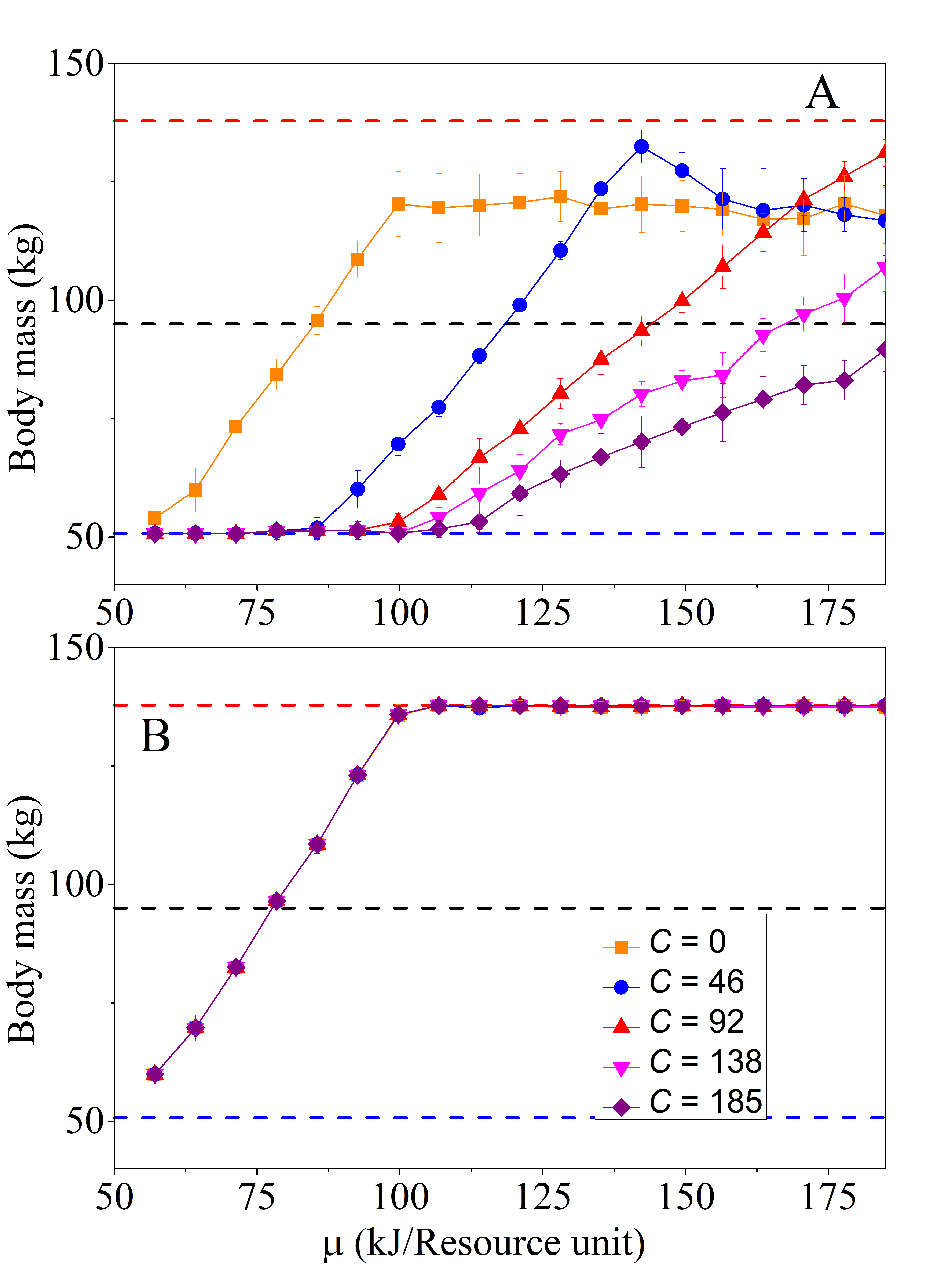}
\caption{Minimum (A) and maximum (B) values of the final mass distribution of males that were able to establish a territory of size greater than 1 (initial condition), in relation with the energy available in the resource, $\mu$, for different values of confrontation cost, $C$. On panel B, all the curves overlap. Each curve is the average of 10 realizations, and we fixed $\alpha = 1$. The dashed lines indicate the average minimum (blue), mean (black) and maximum (red) values of the initial mass distribution, considering a total of 50 males.}
\label{fig:Masa2}
\end{figure}

The effect of including an energetic confrontation cost, $C>0$, is shown in Fig.~\ref{fig:Masa2}, in panels A (the minimum of the final mass distribution) and B (the maximum). Increasing this parameter initially keeps the minimum value of the final mass distribution equal to that of the initial distribution, indicating that the smallest males persist in the system. Larger values of $C$ maintain this minimum even for higher values of $\mu$, a parameter that, as noted above, favors larger males. For sufficiently large $\mu$, with the corresponding value depending on $C$, the minimum mass begins to rise and can eventually exceed the mean of the initial distribution, indicating that the smallest males, although present initially, do not persist in the simulation and are excluded by competitors.
Comparing these curves, some combinations of $\mu$ and $C$ appear to favor larger males, as indicated by regions of $\mu$ in which the curve associated with a higher confrontation cost lies above that corresponding to a lower cost. For example, the blue and red curves, both with $C>0$, exceed the yellow curve with $C=0$ over certain intervals of $\mu$, rather than across the entire range. Overall, however, a clear trend emerges: higher confrontation costs tend to promote the persistence of smaller males.

On the other hand, as shown in Fig.~\ref{fig:Masa2}B, increasing the confrontation cost has no effect on the maximum value of the final mass distribution, since all curves overlap. This indicates that competition costs do not influence the prevalence of larger males. Instead, the advantage observed for smaller males arises from the decision-making process of larger individuals, who choose not to take an unnecessary risk and to avoid confrontation with their less competent conspecifics.

\subsection{Energetic balance}

In line with the economic approach to animal behavior, males are expected to optimize their energetic balances, accounting for the energy gained from territory resources and the costs of border patrolling, in order to maximize their fitness \citep{mitchell2012foraging, cassini2009sociality, flores2020modelling}. In our model, a male achieves a positive energetic balance if $B_i > 0$, as defined in Eq.~(\ref{eq:Balance1}).

We analyzed how the propagation of resources ($\alpha$), the energetic input from resources ($\mu$), and the confrontation cost ($C$) affect the final outcome of the simulation, specifically the fraction of males reaching a positive energetic balance. 

First, examining the $(\alpha,\mu)$ parameter plane (not shown here), we found that the heat map of males with positive balance is essentially identical to the fraction of persistent males shown in Fig.~\ref{fig:AlphaMu}A. This result indicates that, under these conditions, every male that persists in the system maintains a positive energetic balance, which is consistent with expectations for real populations.

\begin{figure}[h]
\centering
\includegraphics[width=0.95\columnwidth]{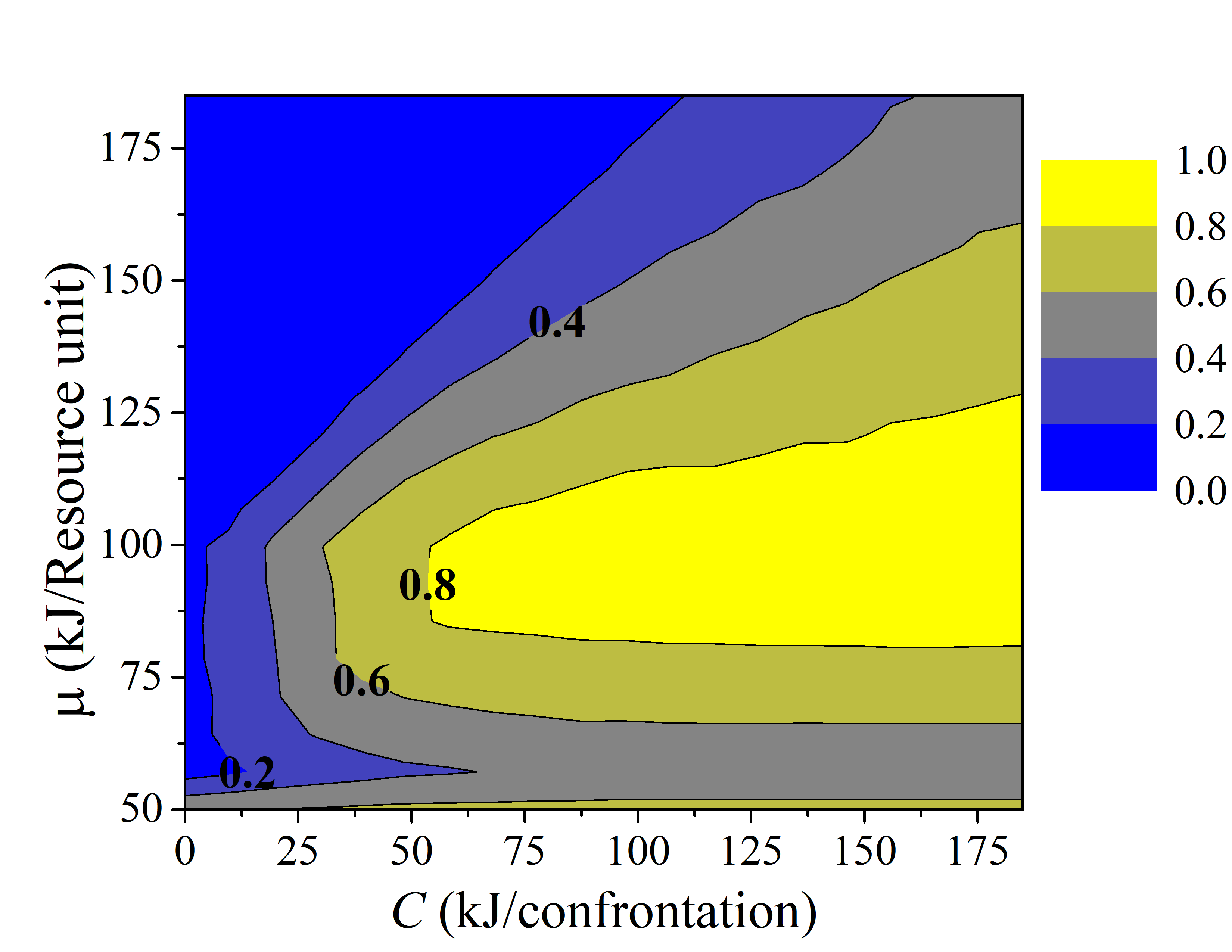}
\caption{Heat map in the $(\mu,C)$ plane showing the fraction of persistent males that achieve a positive energetic balance after 90 iterations. Each pixel represents the average over 20 realizations, with $\alpha = 1$.}
\label{fig:BP}
\end{figure}

More interestingly, analyzing the $(\mu,C)$ plane, we obtained the heat map shown in Fig.~\ref{fig:BP}. As observed, higher confrontation costs generally increase the fraction of males reaching a positive balance.  That is, these are males that could optimize the perimeter-area relationship of their respective territories, securing resources and capable of fulfilling their protective and vigilant role. A minor proportion of males with positive balance is usually due to the action of invading males  over other territories, which increases the complexity of the borders and the costs of patrolling. However, for richer resources (larger $\mu$), higher confrontation costs are required to produce this effect. When $\mu$ is moderate and $C$ is large, almost all males achieve a positive balance, highlighting the interplay between resource abundance and the cost of territorial defense.

\subsection{Territory shape and borders}

\begin{figure}[h]
\centering
\includegraphics[width=1.1\columnwidth]{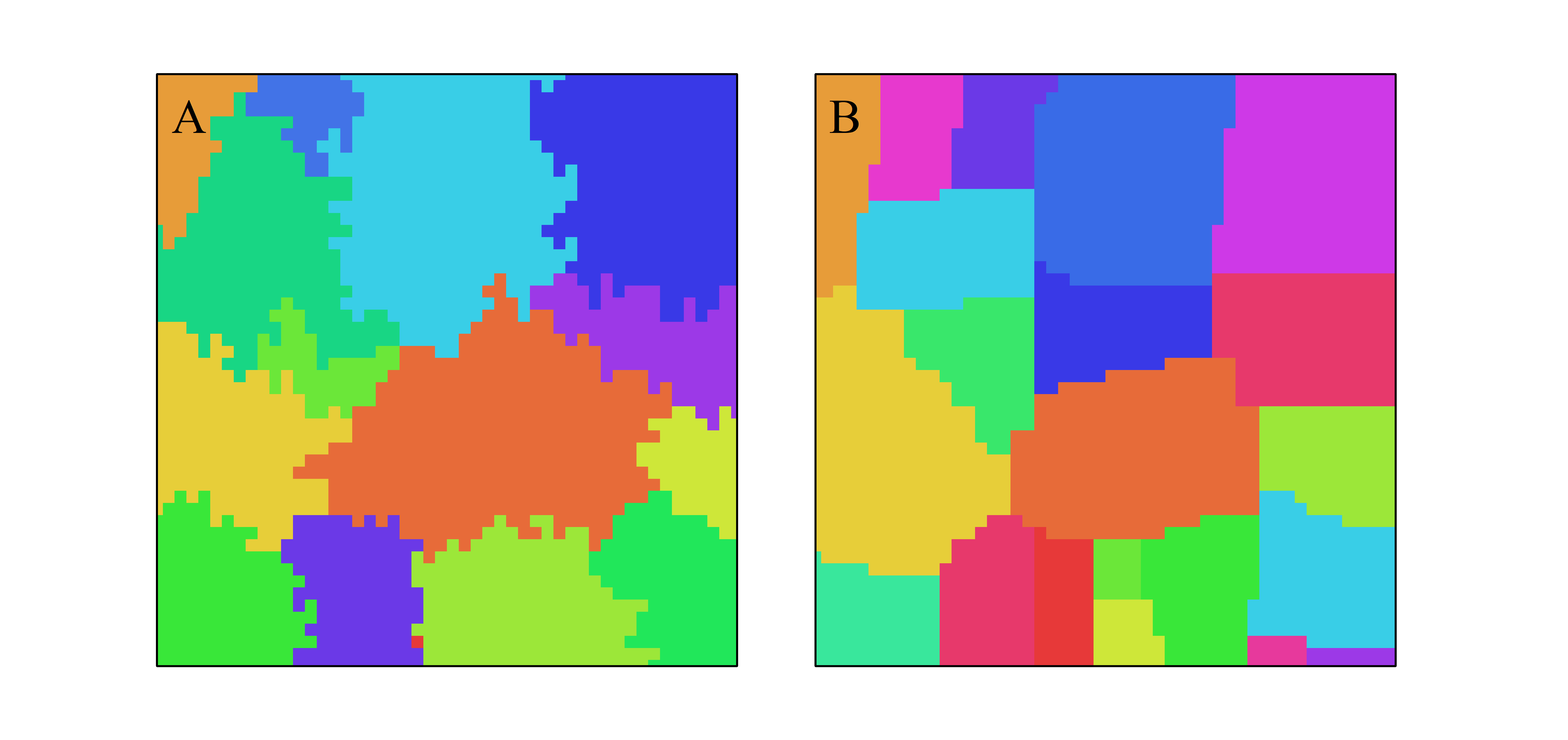}
\caption{Spatial distribution and shape of territories. Each color represents a different male. A) Confrontation cost $C = 0$; B) Confrontation cost $C = 46$. Parameters: $\alpha = 1$, $\mu = 117$.}
\label{fig:Bordes}
\end{figure}

In landscape ecology, the relationship between patch area and perimeter is often used to quantify border effects and infer ecological stability \cite{tiefenbacher2012perspectives}. We applied the same approach to analyze the shape of male territories in our simulations. Examples of territorial configurations are shown in Fig.~\ref{fig:Bordes}.

As seen in Fig.~\ref{fig:Bordes}, when confrontation costs are absent ($C = 0$), territorial borders are intricate, resulting in longer perimeters and higher border effects (this means, a larger effect of each territory over its neighbors due the longer frontiers).

Table~\ref{tab:table} shows the average PAFRAC (see Eq.~\ref{eq:pafrac}) for different confrontation costs $C$. Increasing $C$ reduces the fractal index, indicating simpler, more compact territories with shorter borders. This suggests reduced border effects and increased territorial stability.

\begin{table}[h]
\centering
\begin{tabular}{lcc}
\hline
$C$ & PAFRAC Index & $\sigma$ \\
\hline
0 & 1.20 & 0.03\\
46 & 1.18 & 0.05\\ 
92 & 1.07 & 0.04 \\ 
138 & 1.03 & 0.03 \\ 
185 & 1.02 & 0.04\\ 
\hline
\end{tabular}
\caption{Average and standard deviation of the Perimeter-Area fractal index for territorial areas at the final state, for different confrontation costs $C$, considering 20 realizations.}
\label{tab:table}
\end{table}

\begin{figure}[h]
\centering
\includegraphics[width=1.1\columnwidth]{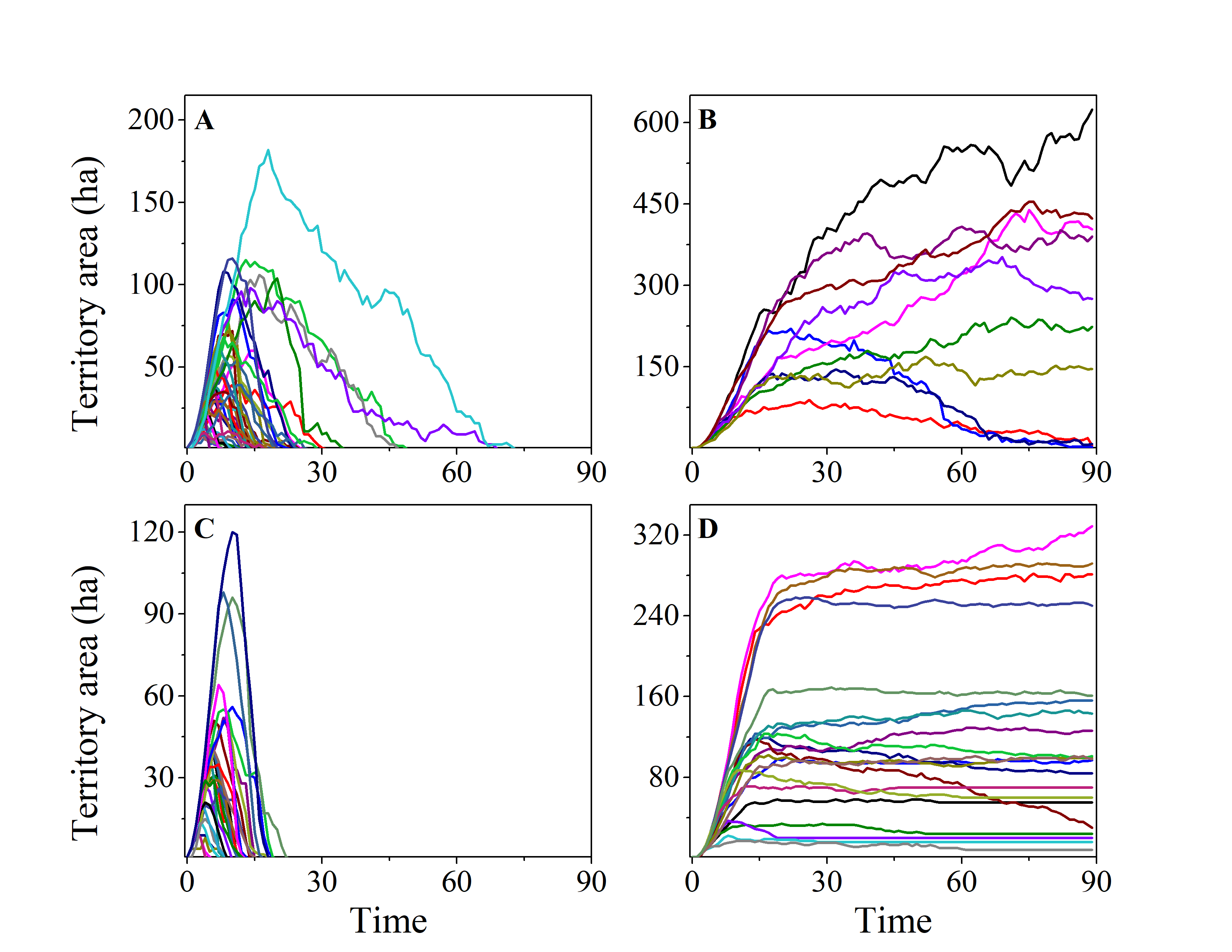}
\caption{Temporal evolution of territorial areas under two confrontation costs: $C=0$ (A, B) and $C=46$ (C, D). Panels A and C show excluded males, while panels B and D show persistent males. Parameters: $\alpha = 1$, $\mu = 117$.}
\label{fig:Areas}
\end{figure}

Figure~\ref{fig:Areas} complements this analysis by showing the temporal evolution of territorial areas under two confrontation regimes ($C = 0$ and $C = 46$). Panels A and C correspond to males that are eventually excluded from the system, whereas panels B and D show persistent males. The comparison reveals that higher confrontation costs lead to more stable territories over time, as indicated by smaller fluctuations and fewer extreme expansions or contractions.

To interpret these dynamics, we examined the body mass of the males whose trajectories are shown in the figure. In the zero-cost scenario (panels A and B), three groups can be distinguished. The first group consists of large males that successfully establish territories, as their size largely determines the outcome of confrontations; these correspond to the trajectories shown in panel B. The second group also comprises large males of similar size, but these individuals are mostly excluded due to stochasticity during confrontations, despite having a considerable probability of winning. The third group consists of smaller males that lose most confrontations due to their lower body mass. These latter two groups correspond to the trajectories observed in panel A.

Considering now the scenario with $C = 46$ (panels C and D), and starting from the same initial mass distribution, we observe that males belonging to the second group in the zero-cost scenario are now able to establish territories, while only the smaller males remain excluded. This interpretation is supported by an analysis of the body mass of the individuals shown in the figure. Specifically, males from the intermediate group identified in the zero-cost case now correspond to the trajectories in panel D, whereas the excluded trajectories in panel C are predominantly associated with smaller males. This occurs because large males avoid confronting others of similar size when the probability of winning is not clearly in their favor, and the associated risk of losing is high; instead, they refrain from invading occupied territories. We therefore conclude that the stable state shown in Fig.~\ref{fig:Areas} emerges from confrontation costs effectively equalizing males, together with a conservative behavioral response to costly conflicts.

\begin{figure}[h]
\centering
\includegraphics[width=1.1\columnwidth]{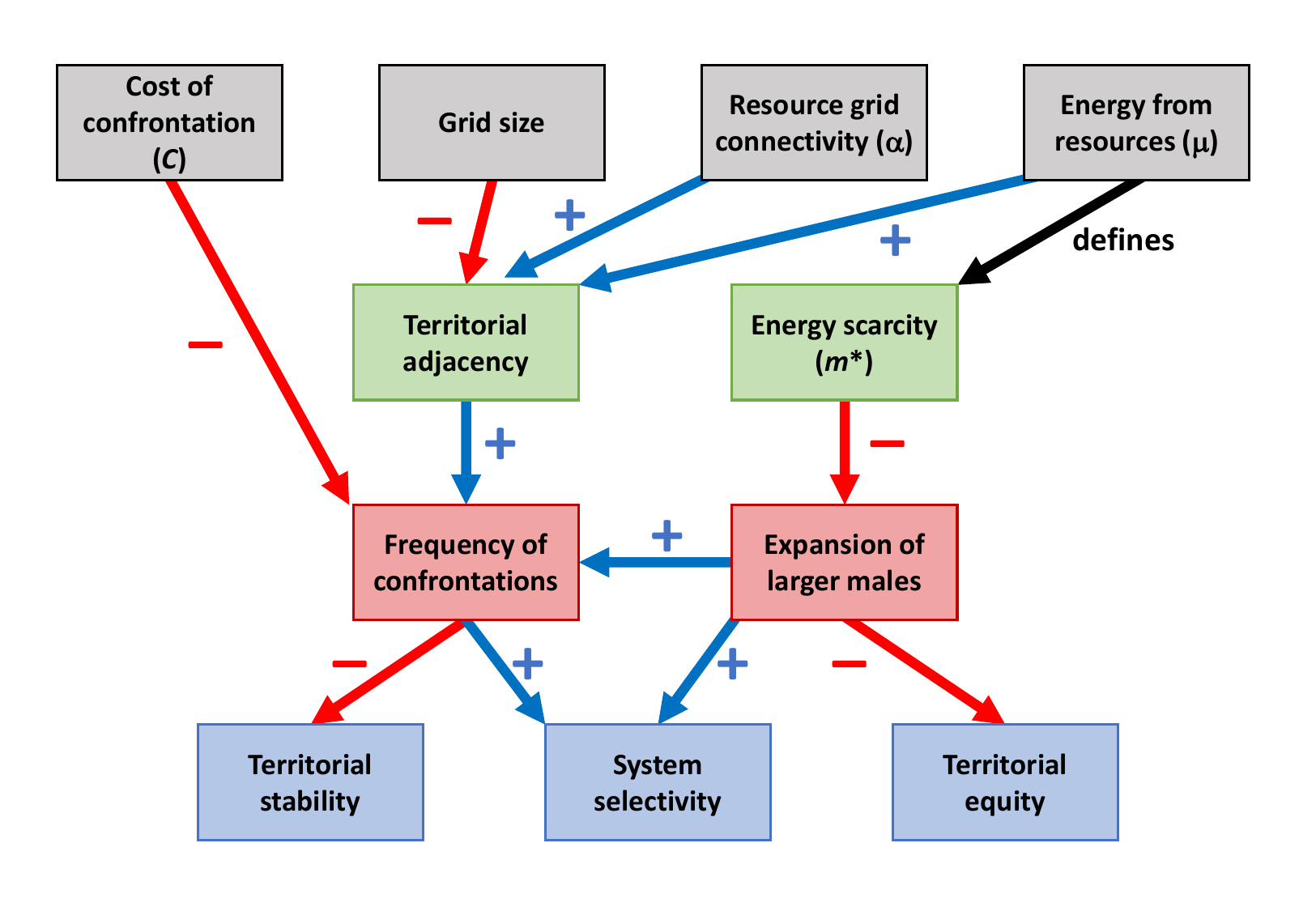}
\caption{Summary of the current results. Grey boxes indicate parameters that are defined before the simulation; green boxes indicate properties that are directly influenced by parameters; red boxes indicate processes that take place during the simulation, affected by the properties; and blue boxes indicate features of the final state, collectively defined by the process. The red arrows indicate negative correlation; the blue arrows indicate positive correlation.}

\label{fig:Esque}
\end{figure}

\section{Discussion}

Our results show that the interplay between energetic constraints, spatial connectivity, and individual competition strongly shapes territorial organization in systems of competing males. Using an individual-based spatial model, we show how these factors jointly regulate the persistence of males, the distribution of territory sizes, and the emergence of unequal territorial structures (see a schematic representation of our results in Fig.~\ref{fig:Esque}). Although the model considers a single species and only intraspecific competition, the results reveal mechanisms that resemble patterns predicted by broader ecological theories of competition and coexistence.

In classical Lotka-Volterra competition models and in Modern Coexistence Theory, coexistence among competitors is expected when intraspecific competition is stronger than interspecific competition. However, these approaches are typically phenomenological and summarize competitive interactions through simple terms such as $1/K_i$ (where $K_i$ is the system carrying capacity for each population $i$) \citep{ayala1973competition, anazawa2012interspecific, chesson2012species, johnson2022resolving}. In contrast, our individual-based model generates a comparable stabilizing effect through explicit energetic confrontation costs between individuals. These costs limit territorial expansion and prevent the progressive exclusion of competitors, ultimately stabilizing both the number of males and the distribution of territories. Moreover, the model allows us to explore how these competitive costs interact with other ecological factors, such as resource distribution and spatial connectivity.

Landscape connectivity emerges in our simulations as a key factor shaping the outcome of territorial competition. Connectivity, understood here as the structural continuity between resource patches, has long been recognized as an important determinant of ecosystem functioning and species persistence \citep{kindlmann2008connectivity, correa2016habitat}. In our model, connectivity is controlled by the seeds propagation parameter $\alpha$, which determines how resources spread across the matrix. Different levels of connectivity generate markedly different territorial outcomes: highly connected matrices promote strong competition and the exclusion of many males, whereas fragmented resource distributions allow more individuals to maintain territories. This shows that simple changes in spatial structure can produce non-linear effects on population organization.

From this perspective, matrix connectivity may play a role similar to that proposed by disturbance-based coexistence mechanisms. Reduced connectivity can create spatial refuges that limit direct interactions between dominant and subordinate individuals, allowing smaller or less competitive males to maintain territories. In this sense, spatial fragmentation can operate analogously to the Intermediate Disturbance Hypothesis at the intraspecific level, preventing the monopolization of resources by a few dominant individuals \citep{Connell1978, raerinne2015exclusions}. However, this mechanism must be interpreted in conjunction with energetic constraints, since body size strongly affects locomotion costs according to the allometric relationship described by Taylor and Heglund \citep{taylor1982energetics}.

The model also reveals interesting parallels with systems studied in econophysics. In simulations without confrontation costs, the number of males declines continuously while the distribution of territorial areas converges toward a stable configuration. Similar dynamics have been observed in economic exchange models, where interactions among agents lead to stable but highly unequal wealth distributions \citep{nener2021optimal, dias2024effectiveness, giordano2025limiting}. Such analogies contribute to a growing body of research exploring conceptual links between ecological and economic systems, particularly in terms of resource allocation and competition among agents \citep{rapport1977economic, bloom1985resource, tisdell2004economic}.

Territorial organization also has important implications for population structure and reproductive dynamics. In wild camelids, territorial males defend resource-rich areas in order to attract and retain females. Consequently, variation in territory size may translate into differences in reproductive success, survival, and ultimately genetic contribution to future generations \citep{franklin1983contrasting, young2004activity, adams2001approaches, owen1977territoriality}. Our results suggest that highly connected and energetically rich environments tend to concentrate resources into fewer territories, potentially reducing genetic diversity by increasing reproductive inequality among males. Conversely, stronger confrontation costs tend to favor the persistence of more males with positive balances, and therefore stable territories, and a more distributed reproductive structure.

These findings complement previous theoretical work on the social organization of wild camelids. In a previous study we had analyzed how female distribution emerges from the interaction between territory quality, male performance, and the presence of bachelor males in the system \citep{GONZALEZ2026111542}. Combining both approaches suggests that highly productive and well-connected environments may generate social contexts where many males fail to secure territories and instead adopt harassment strategies. Under these conditions, females may prefer to form larger groups despite the potential costs associated with resource competition within groups. Such dynamics can further increase inequality in the distribution of females among territories, adding another layer of complexity to the system.

Our modeling approach is also related to economic models of animal territoriality, which interpret territorial behavior as the outcome of cost–benefit optimization strategies. Many of these models assume that individuals adjust their behavior to maximize energetic benefits relative to costs affecting fitness \citep{mitchell2012foraging, dill1978energy, adams2001approaches, hixon1980food, hixon1987territory}. However, most of these models are deterministic, non-spatial, and focus on the decision-making process of a single territorial resident. In contrast, our model explicitly incorporates spatial structure and the simultaneous interaction of multiple competing males, allowing the emergent territorial configuration to arise from decentralized individual decisions.

Another important component of our model is the explicit incorporation of allometric constraints. Locomotion costs associated with territorial patrol depend strongly on body mass and biomechanical limitations. These energetic costs ultimately influence the size and stability of territories, linking individual physiological traits with population-level spatial organization \citep{sibly2007effects,brown2004toward, cloyed2021allometry, bowyer2020evolution}. In this sense, the energetic balance framework provides a mechanistic bridge between individual traits and emergent social structures.

Finally, it is important to compare our model outcomes with empirical observations. In wild camelids, typical territory sizes range between approximately 13 and 46 hectares \citep{koford1961ecology, young2004activity, arzamendia2018social, smith2020and}. Some of our simulations, particularly those with very high connectivity, generate territories larger than these values. In real systems, several ecological factors (that are not included in the model) may constrain territorial expansion, including selective grazing, habitat visibility, predation risk, and interactions with other members of the family group. Nevertheless, large territories have been reported in some guanaco populations, reaching between 200 and 900 hectares in certain contexts \citep{marino2008vigilance}.

Regarding energetic inputs, the main grasses consumed by wild camelids belong to genera such as \textit{Stipa}, \textit{Poa} and \textit{Panicum}. Considering the approximate caloric values of these species, the total energy available in natural systems may exceed the values explored in our simulations. However, real males must share these resources with females and offspring and also incur additional energetic costs associated with social monitoring, territorial defense, and interactions with bachelor males \citep{borgnia2010foraging, puig1997diet, puig2014food, kelrick1986native, hadley1963productivity, sobol2022determination, marino2012indirect, marino2014ecological, GONZALEZ2026111542}. Incorporating these processes into future models could help bridge the gap between theoretical predictions and empirical observations.

In summary, our study illustrates how territorial organization can emerge from the interaction between energetic constraints, spatial structure and individual competition. By explicitly linking allometric scaling to competitive dynamics, the framework provides a mechanistic perspective connecting individual physiology to population-level patterns. Beyond camelids, these results suggest that incorporating energetic and spatial constraints into models of territoriality may be essential for understanding the diversity and stability of social systems in vertebrates.

\section{Data availability}
The computational code generated during the current study are available in the Mendeley Data repository, \url{https://doi.org/10.17632/byc5w9jxjs.1}.

\section{Acknowledgements}
This research was supported by Consejo Nacional de Investigaciones Científicas y Técnicas (CONICET, PIP 112-2022-0100160 CO).

\bibliographystyle{elsarticle-harv}
\bibliography{ref}

\clearpage
\appendix

\renewcommand{\thesection}{S\arabic{section}}
\renewcommand{\thefigure}{S\arabic{figure}}
\renewcommand{\thetable}{S\arabic{table}}

\setcounter{section}{0}
\setcounter{figure}{0}
\setcounter{table}{0}

\makeatletter
\let\@title\@empty     
\renewcommand{\MaketitleBox}{
  \resetTitleCounters
  \def\baselinestretch{1}%
  \begin{center}
    \def\baselinestretch{1}%
    \Large \@title \par       
    \vskip 18pt
    \normalsize\elsauthors \par 
    \vskip 10pt
    \footnotesize \itshape \elsaddress \par 
  \end{center}
  \vskip 12pt
}
\makeatother

\title{Supplementary Material: \\
Energy, space and competition: A model of territorial organization in camelids}
\maketitle

\section{Generation of the resource matrix}
In this section, we describe the procedure used to generate the matrix representing the spatial distribution of resources in the simulations.

To construct the resource matrix, we implemented a custom function in two steps: (i) sowing and (ii) propagation. In the sowing step, $N$ grid cells are randomly selected as sowing points. Each of these cells is assigned a value of 1, representing the maximum resource capacity.

In the propagation step, the remaining cells receive contributions from the sowing points in an amount that depends on their Euclidean distance to each sowing point. The value of cell $(i,j)$ is computed as:
\[
M_{i,j} = \sum_{z=1}^{N} M_z D_z^{\alpha} ,
\]
where $M$ is the resource matrix, $D_z$ is the distance from cell $(i,j)$ to sowing point $z$, and $M_z$ is the value at sowing point $z$. The parameter $\alpha$ controls the degree of resource propagation.

Because resource values represent the fraction of the maximum capacity of a cell, we impose an upper bound of 1 using the following piecewise definition:

\[
M_{i,j} =
\begin{cases}
\displaystyle \sum_{z=1}^{N} M_z D_z^{\alpha}, & \text{if } \sum_{z=1}^{N} M_z D_z^{\alpha} \le 1, \\[6pt]
1, & \text{if } \sum_{z=1}^{N} M_z D_z^{\alpha} > 1 .
\end{cases}
\]

We define the resource grid in terms of normalized cell capacity rather than specific resource units (e.g., vegetation bio\-mass or water mass). This choice avoids unnecessary distinctions for the purposes of the model and allows the energetic gain from resources to be quantified separately through the parameter $\mu$ (see the corresponding sections of the main text).

\begin{figure}[h]
\centering
\includegraphics[width=\columnwidth]{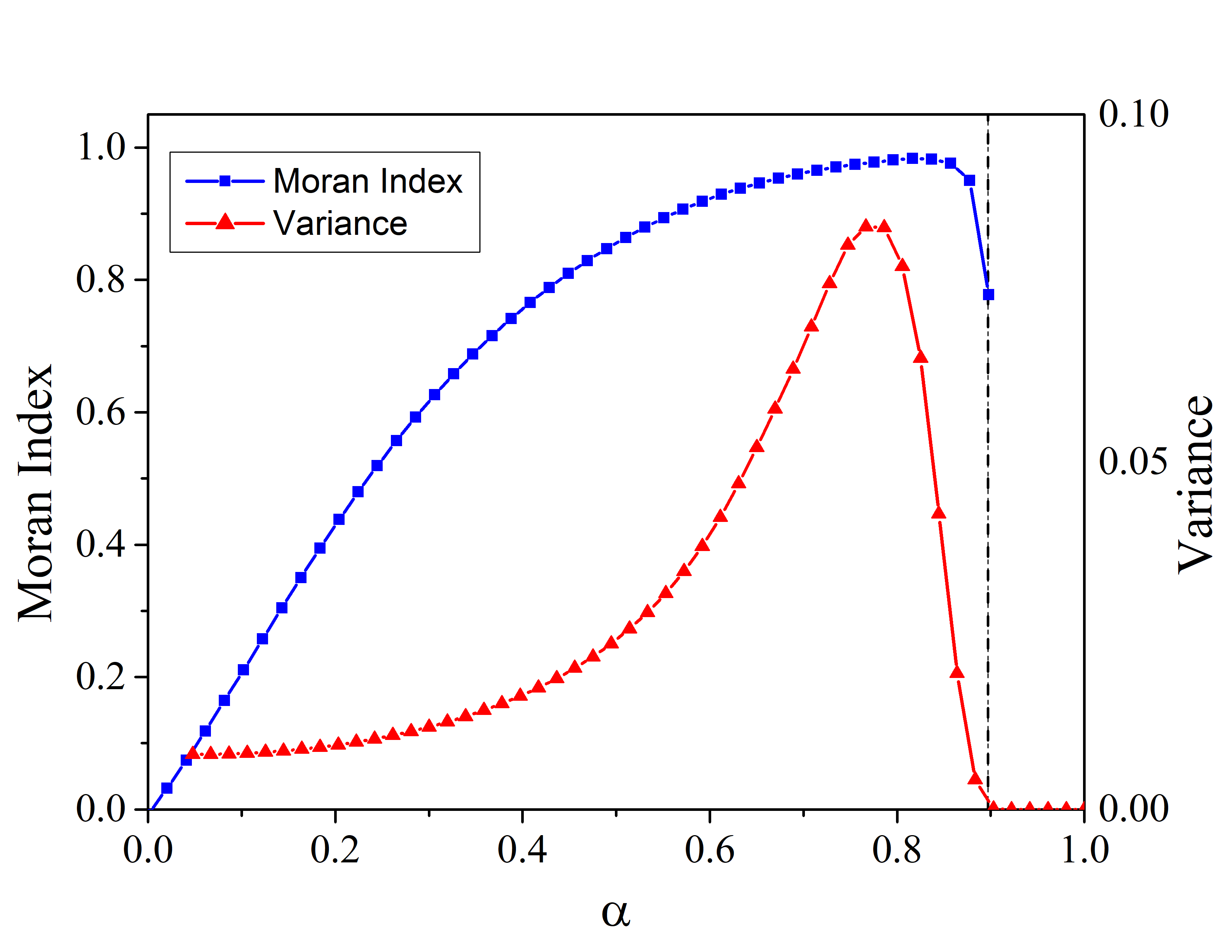}
\caption{Spatial autocorrelation and variance of the generated resource matrix as a function of the propagation parameter $\alpha$. }
\label{fig:Siembra}
\end{figure}
Although alternative procedures could be used to generate the resource matrix, this approach is particularly convenient because the parameter $\alpha$ allows us to control the degree of spatial autocorrelation in the resource distribution. For small values of $\alpha$, resources remain concentrated around the sowing points, generating heterogeneous landscapes. As $\alpha$ increases, resources disperse across the grid and the matrix gradually becomes more homogeneous.

To quantify these patterns, we analyzed the spatial autocorrelation of the matrix using the Moran index \citep{moran1950notes,li2007beyond}:
\[
I = \frac{N}{W}\,
\frac{\sum_{i,j=1}^{N} w_{ij}(x_i-\bar{x})(x_j-\bar{x})}
{\sum_{i=1}^{N}(x_i-\bar{x})^2},
\]
where $x_i$ is the value of the variable at the location $i$, $\bar{x}$ is its mean value, $w_{ij}$ indicates the spatial weight, being equal 1 if the locations $i$ and $j$ are adjacent, or being equal 0 if they are not, and $W$ being the total sum of all elements $w_{ij}$. This index ranges from $-1$ (complete negative spatial correlation) to $1$ (complete positive spatial correlation), and incorporates the overall variance of the matrix through the denominator. As shown in Fig.~\ref{fig:Siembra}, increasing $\alpha$ initially produces a rise in spatial autocorrelation, while the variance changes only moderately. For sufficiently large $\alpha$, the matrix becomes nearly homogeneous, the variance approaches zero, and the Moran index becomes undetermined.

\begin{figure}[t]
\centering
\includegraphics[width=\columnwidth]{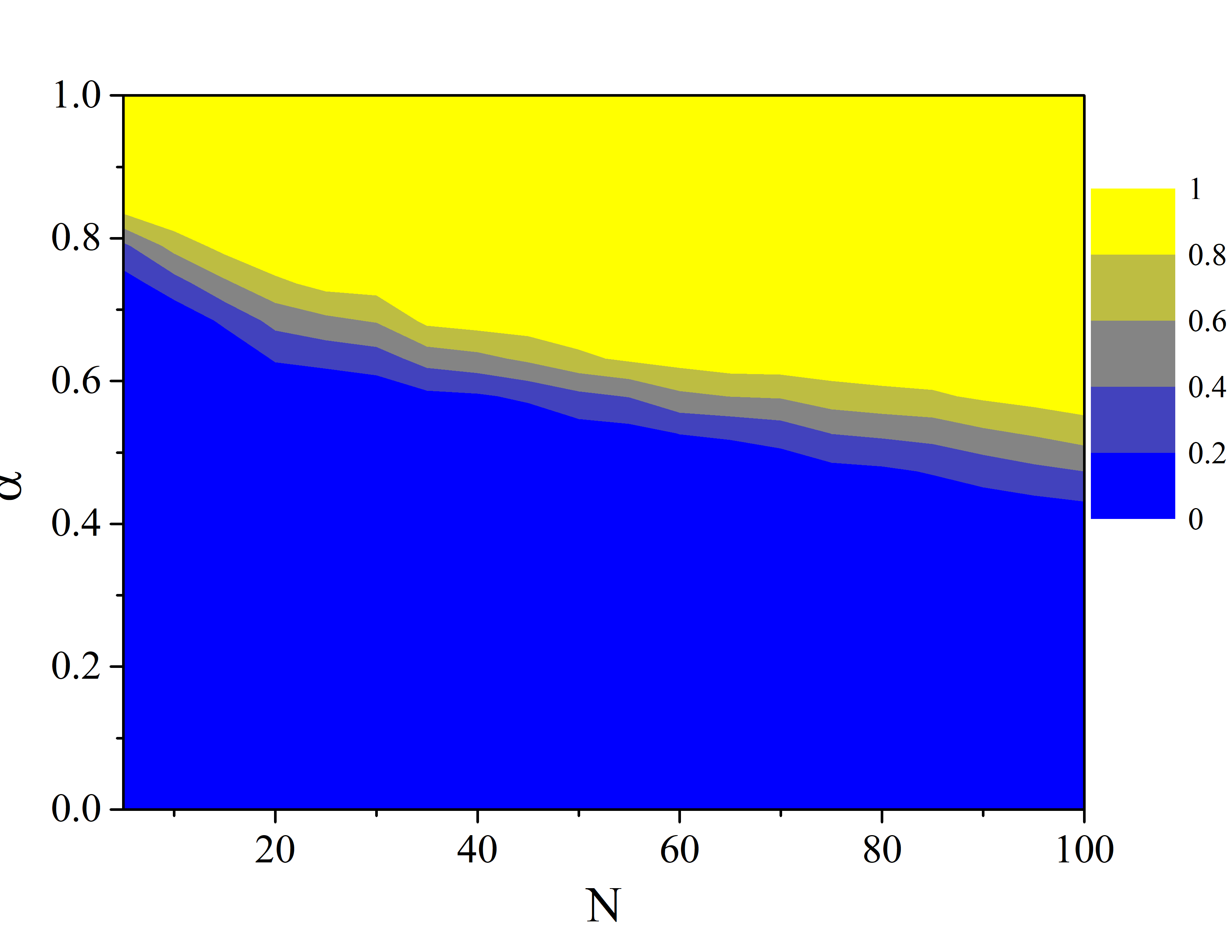}
\caption{Percolation of the resource matrix. The colors correspond to the fraction of rows and columns belonging to the percolating cluster as a function of the propagation parameter, $\alpha$, and the number of sowing points, $N$.}
\label{fig:Perco}
\end{figure}

Finally, this method also allows us to identify when the resource matrix reaches percolation. In Fig.~\ref{fig:Perco}, we show the fraction of rows and columns belonging to the percolating cluster (relative to the total number of rows and columns) as a function of the propagation parameter $\alpha$ and the number of sowing points $N$.

\section{Males conflict}

\begin{figure}[t]
\centering
\includegraphics[width=\columnwidth]{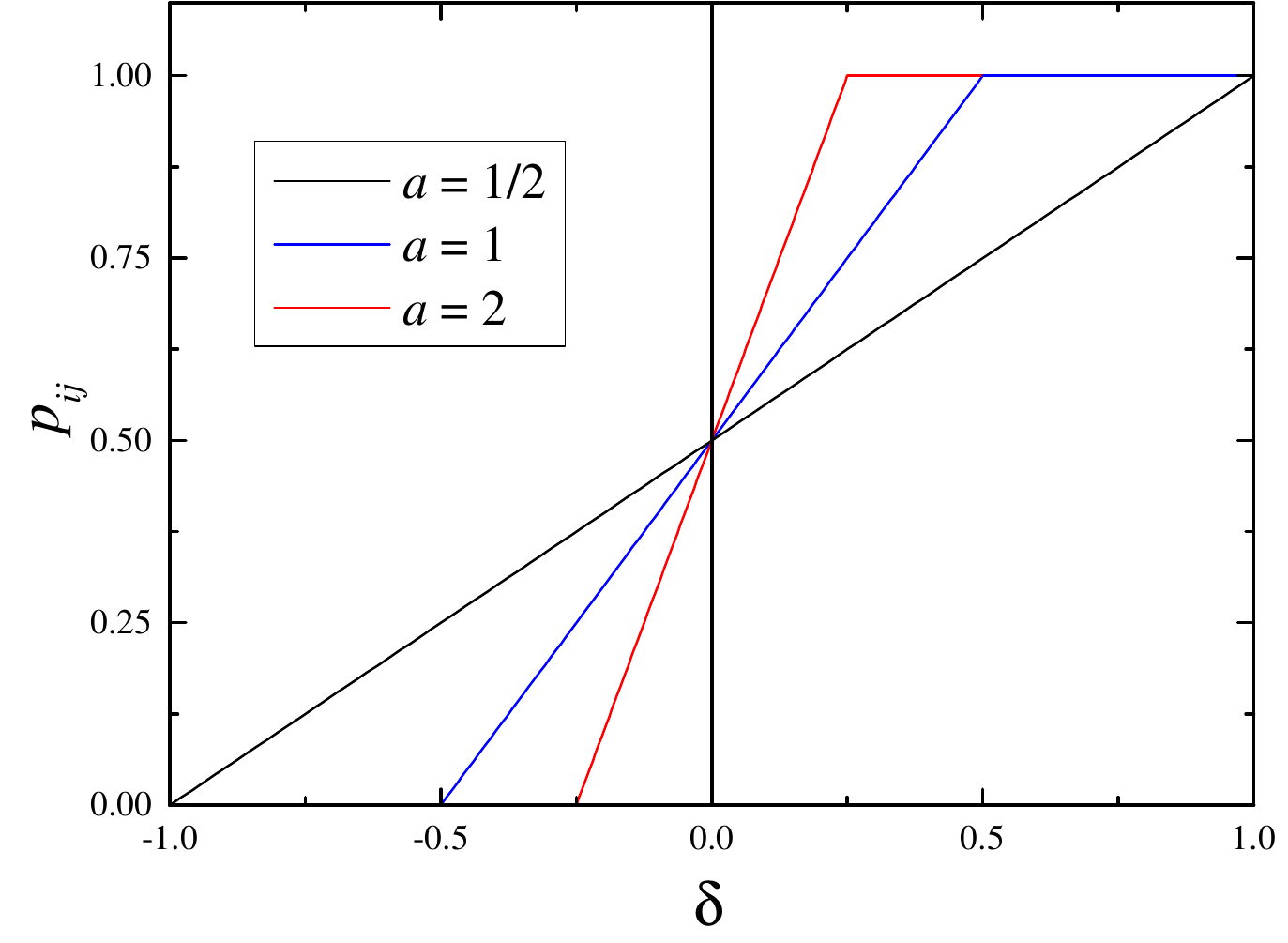}
\caption{Probability that male $i$ wins a conflict against $j$, as a function of the relative mass difference. Black: masses in $(0,\infty)$. Blue: heaviest one is at most 3 times the lightest one (this is the case used in our simulations). Red: heaviest is at most $1.67$ times the lightest one.}
\label{fig:pij}
\end{figure}

Body mass is arguably the most relevant trait in the conflict between males, deeply intertwined with other factors such as social status, age, and behavior
\citep{wilson1985male}. 
In our model, we chose the relative mass difference between an invading male, $i$, and the incumbent of a territory, $j$:
\[
\delta = \frac{m_i-m_j}{m_i+m_j},
\]
as the parameter that controls both the decision of $i$ to initiate an attack, and the probability that it wins it. We can see that, when $m_i\gg m_j$, $\delta\to 1$, and conversely, when $m_i\ll m_j$, $\delta\to -1$. So, in principle, one needs to define the probability that $i$ attacks (and also the probability of winning), $p_{ij}$, with compact support $\delta\in(-1,1)$. Moreover, since the larger male has the upper hand, one should have:
\[
p_{ij} \to \begin{cases}
           0& \text{when }\delta\to -1,\\
		   1& \text{when }\delta\to 1.
		   \end{cases}
\]
Besides, the probability should be $1/2$ when the masses are equal (since we are assuming that body mass is the only relevant trait). A possible choice, then, is to define $p_{ij}$ as a linear function between these extremes (shown in Fig.~\ref{fig:pij}, in black). However, this assumes, implicitly, that the masses can have any value between 0 and infinity. This is not a reasonable assumption for any real males population. Rather, the ratio between the heaviest and the lightest males would be a factor not much larger than one. In the case of guanacos, adult masses are documented between 50 and 140 kg, so that the factor is $\sim 3$, and one can take $\delta\in(-1/2,1/2)$, interpolating $p_{ij}$ linearly between these extremes. This was our choice in the model of male guanacos, and it is shown in Fig.~\ref{fig:pij} in blue. 

Of course, if a species had a different distribution of body masses, the same idea can be applied. The probablity can be defined as $p_{ij}= 1/2+a\,\delta$, with an adequate value of $a$. If the distribution is very narrow, then $a$ would be typically large (in our case, $a=1$ for a mass ratio of 3). Conversely, if the range of possible masses is wider, the value of $a$ would be smaller than 1, and tending to $1/2$ when it goes to infinity (the black curve). The case $a=2$, corresponding to a narrow distribution of masses, with  an extreme mass ratio of $5/3\approx 1.7$, is shown in red in the figure. 

Nonlinear interpolations between the extremes are possible, but would require additional parameters in the definition, without adding any qualitative advantage. Our choice is parameter-free, since it depends only on the extreme values of the distribution of body masses, which in principle can be directly observed in the field.

\section{Mass threshold}

In the main text we introduced a mass threshold that determines whether a male can expand its territory while maintaining a non-negative energetic balance (Eq.~10). This threshold corresponds to the purple curve shown in Fig.~5, which separates the region of body masses compatible with territorial expansion from the region where expansion becomes energetically unfeasible. In this section we provide the derivation of that threshold under a simplified configuration of the system.

As a first approximation, we analyze the model in a homogeneous environment in which all cells of the resource matrix have maximum capacity ($M_{i,j} = 1,~\forall i,j$). Since the system is governed by the energetic balance equation (Eq.~6), we evaluate the energetic condition required for a male to incorporate its first neighboring cell.

\begin{figure}[t]
\centering
\includegraphics[width=\columnwidth]{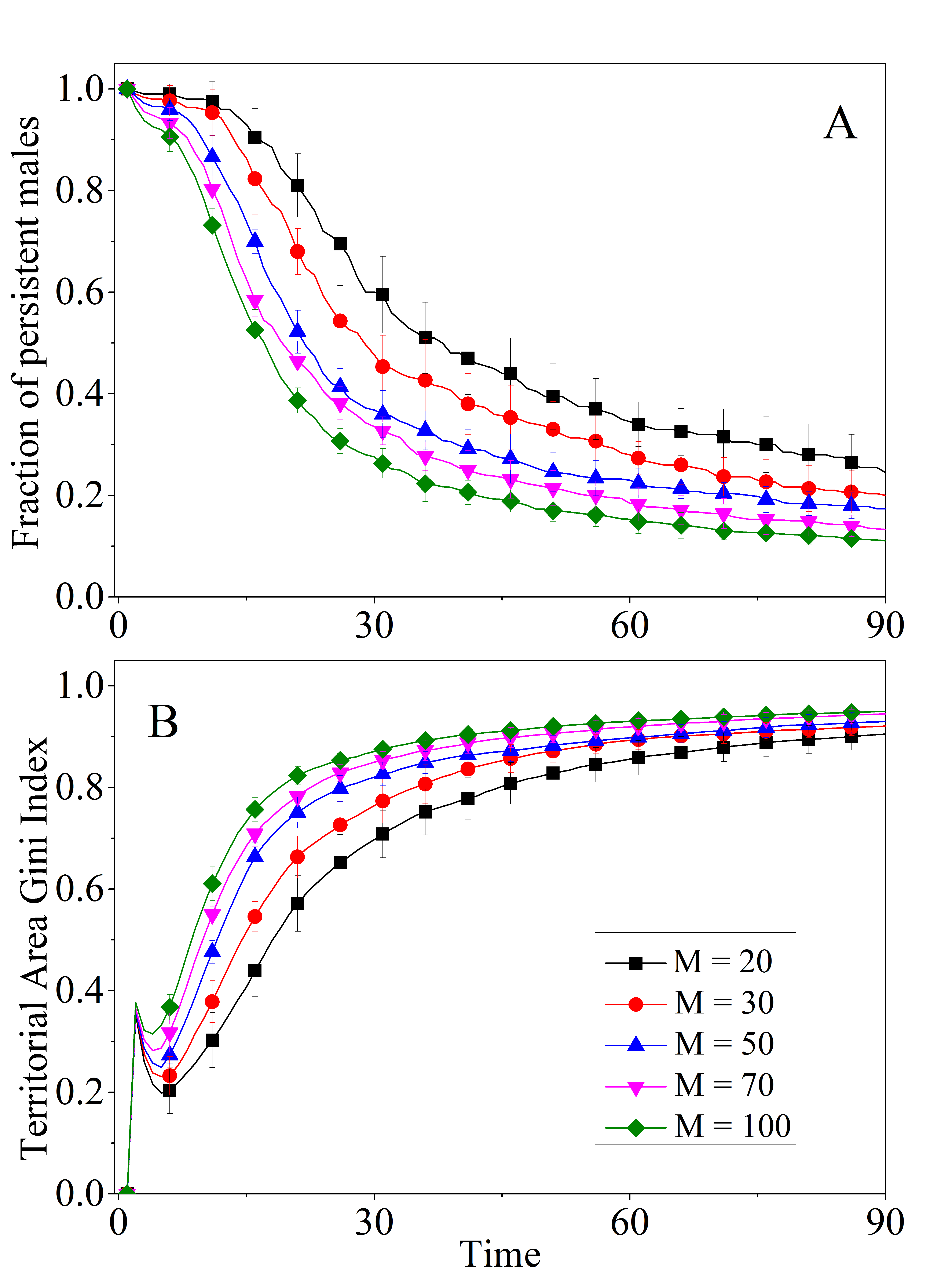}
\caption{Time series of the fraction of persistent males (A) and the Gini index of territorial areas (B) for different total numbers of males in the system, as indicated in the figure.}
\label{fig:NEffect}
\end{figure}

As described in the model, territorial expansion occurs by incorporating nearest-neighbor cells. Therefore, when a male evaluates the incorporation of its first neighboring cell, the mar\-gi\-nal resource gain corresponds to the resource value of that cell. Under the homogeneous assumption this implies $\Delta R = 1$. At the same time, incorporating this cell increases the perimeter of the territory by $\Delta P = 2$. Under these conditions, the energetic balance becomes:
\[
\mu \Delta R - 100\, m_i E(m_i)\Delta P > 0
\]
\[
\Rightarrow
\mu - 200\, m_i E(m_i) > 0
\]
\[
\Rightarrow
\mu > 200\, m_i E(m_i).
\]

To further develop this expression, we use the locomotion cost equation proposed by Taylor and Heglund (1982):
\[
E(m,V) = 10.7\,m^{-0.316}V + 6.03\,m^{-0.303}.
\]

\begin{figure}[t]
\centering
\includegraphics[width=\columnwidth]{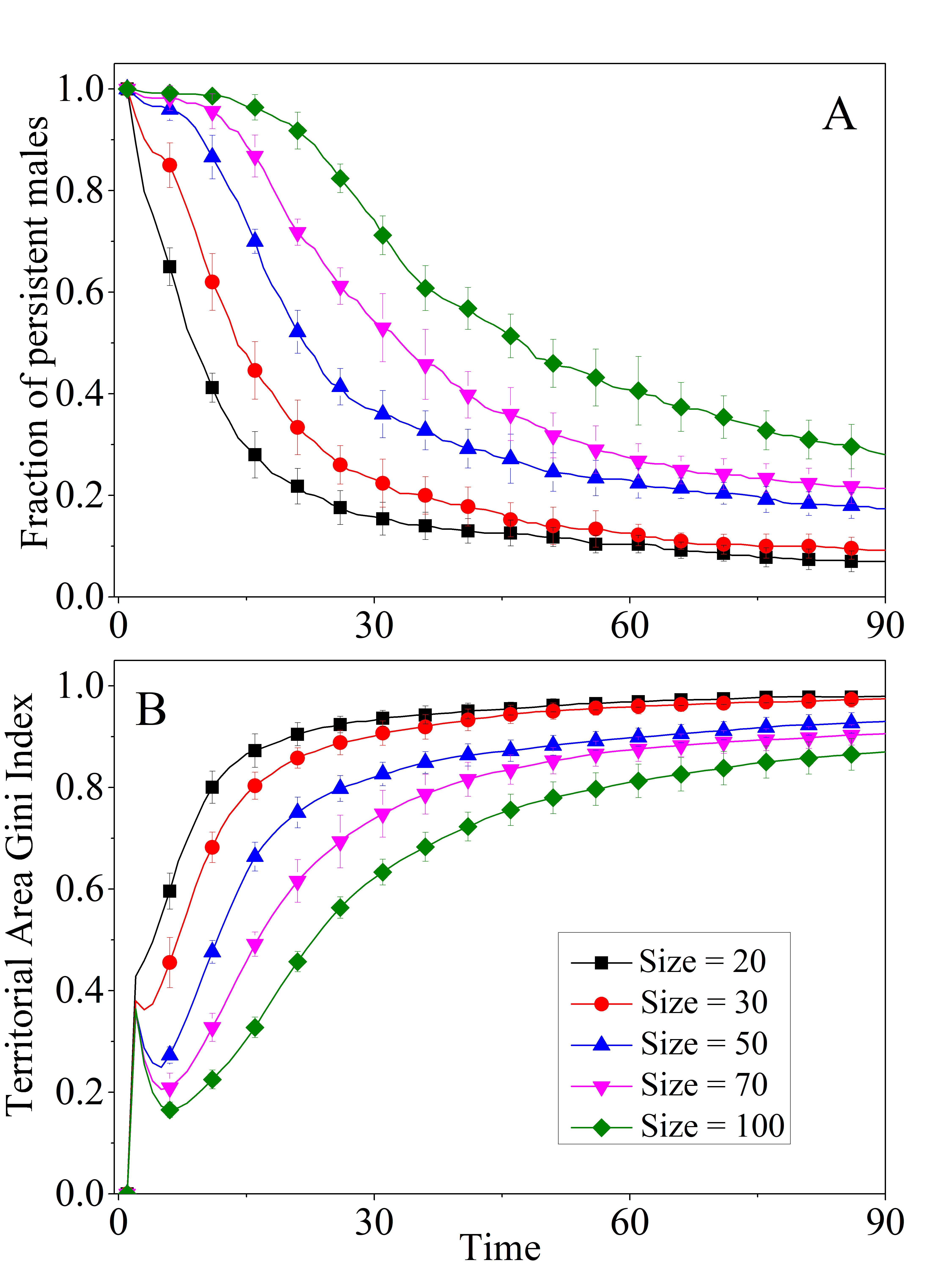}
\caption{Time series of the fraction of persistent males (A) and the Gini index of territorial areas (B) for different grid sizes.}
\label{fig:SizeEffect}
\end{figure}

Since the exponents of body mass in both terms are very similar, we approximate them by their mean value. This simplification introduces a negligible error while allowing us to obtain a closed-form scaling relation between body mass and the energetic threshold:
\[
E(m,V) \sim 10.7\, m^{-0.309}V + 6.03\, m^{-0.309}.
\]
In the model, we assumed a constant patrol velocity $V = 1\,\mathrm{m\,s^{-1}}$, which simplifies the expression to:
\[
E(m,V) \sim 10.7\, m^{-0.309} + 6.03\, m^{-0.309} = 16.73\, m^{-0.309}.
\]
Substituting this expression into the energetic balance condition yields:
\[
200\, m_iE(m_i) = 3346\,m_i^{0.691}.
\]

From this relation we obtain a mass threshold that determines how large a male can be while still being able to incorporate its first neighboring cell without compromising its energetic balance:
\[
m < \psi \mu^{1.45},
\]
where $\psi = 1/3346^{1.45}\,\mathrm{J^{-1.45}\,kg}$ is a numerical coefficient that incorporates the spatial scales defining the territory and the allometric relationship between body mass and energetic costs. This threshold therefore defines the maximum body mass compatible with territorial expansion under the energetic constraints of the model.

\section{Effects of population and system size}

As a complement to the results presented in the main text, we analyzed how population size and grid dimension influence the dynamics of the model. To isolate these effects, we used a fully connected, homogeneous resource matrix ($\alpha = 1$) and set the confrontation cost to $C = 0$.

As observed in Fig.~\ref{fig:NEffect}A, systems with larger populations within the same grid size (i.e., higher population density) eliminate less competitive males more rapidly, resulting in a smaller fraction of persistent males. At the same time, the Gini index of territorial areas increases, indicating a more uneven distribution of territories (Fig.~\ref{fig:NEffect}B).

Conversely, increasing the grid size produces the opposite effect. As shown in Fig.~\ref{fig:SizeEffect}, larger grids allow more males to persist in the system and reduce the Gini index, reflecting a more even distribution of territorial areas.

\end{document}